# Restoring Consistency with Poynting's Theorem in Numerical MHD and RMHD

**By**

**Dinshaw S. Balsara[1,2] and Deepak Bhoriya[3]**

**[1]Physics and Astronomy Department, [2]ACMS Department, University of Notre Dame (dbalsara@nd.edu)**

**[3]Department of Mathematics, Birla Institute of Technology & Science, Pilani**

**Abstract**

For perfectly conducting plasmas, Poynting's theorem states that the time rate of change of electromagnetic energy density equals the negative of the divergence of the Poynting flux. To the best of our knowledge, this involutive theorem has never been mimetically incorporated in codes. We show that there is an inconsistency in MHD and RMHD codes that evolve face-centered magnetic fields in globally-divergence-free fashion. This is because the face-centered flux of energy density, which incorporates a Poynting vector, is evaluated using one-dimensional Riemann solvers that are applied at face-centers. However, the face-centered magnetic fields are evolved according to edge-centered electric fields that come from two-dimensional Riemann solvers that are applied to the edges of a zone. While edge-centered electric fields are used to evolve the magnetic field, they have no influence on the face-centered Poynting flux. For situations where the magnetic energy is much larger than the thermal energy, this causes a negative pressure to be evaluated at the zone-centers. This paper restores consistency by using edge-centered variables to evaluate a face-centered Poynting flux. A simple accuracy-preserving correction is made to the face-centered flux of energy density. We show that it helps considerably in simulations that have very strong magnetic fields.

## I) Introduction

Astrophysical numerical magnetohydrodynamics (MHD) and numerical relativistic magnetohydrodynamics (RMHD) play an important role in modern day astrophysics and space physics simulations. In addition to the hydrodynamic equations, the system requires the incorporation of Faraday's law, which posits the globally divergence-free evolution of face-centered magnetic fields using edge-centered electric fields. Ever since the early work of Yee (1966), it has been realized that mimetic formulations work better than those that do not respect the involution inherent in Faraday's law. This drives home the point that mimetically incorporating as many involutive constraints as possible always seems to bring a benefit. Early works (Brecht *et al*. 1981, DeVore 1991, Evans and Hawley 1989) were based on artificial viscosity formulations. Later works (Brio and Wu 1988, Dai and Woodward 1994a,b, Zachary *et al*. 1994, Powell 1994, Ryu and Jones 1995, Roe and Balsara 1996, Cargo and Gallice 1997, Balsara 1998a,b, Ryu et al. 1998, Dai and Woodward 1998, Balsara and Spicer 1999a,b, Toth 2000, Dedner *et al*. 2002, Balsara 2001a, 2004, 2009, 2010, 2012a, 2014, Balsara and Kim 2004, Londrillo and DelZanna 2004, Gurski 2004, Li 2005, Gardiner and Stone 2005, Miyoshi and Kusano 2007, Bouchut et al. 2007, 2010, Lee and Deane 2009, Waagan 2009, Klingenberg and Waagan 2010, Waagan *et al*. 2011, Lee 2013, Balsara and Dumbser 2015a,b, Dumbser and Balsara 2016, Felker and Stone 2018, Balsara and Nkonga 2017, Balsara and Käppeli 2017, Wu 2018, Dumbser *et al*. 2019, Wu *et al*. 2018, 2019, 2023a,b, Minoshima *et al*. 2020, 2021, Seo and Ryu 2023, Fambri *et al*. 2023, 2025 Liu *et al*. 2025, Balsara *et al*. 2019, 2021, 2023, 2025a,b, 2026a,b,c, Wu and Shu 2026, Yan et al. 2026, Zhang *et al*. 2026, Wang *et al*. 2025, 2026, Tomida *et al*. 2026) are based on higher order Godunov methods for MHD. A rich literature has also developed for RMHD (Komissarov 1999, Balsara 2001b, Gammie *et al*. 2003, Noble *et al*. 2006, Mignone *et al*. 2006, 2009, Giacomazzo and Rezzolla 2006, 2007, DelZanna *et al*. 2007, Honkkila and Janhunen 2007, Tchekhovskoy *et al*. 2007, Anton *et al*. 2010, McKinney *et al*. 2014, Kim and Balsara 2015, Etienne *et al*. 2015, Balsara and Kim 2016, Cai *et al*. 2025, Teukolsky 2026, Balsara *et al*. 2025a,b, 2026c). The citations in the previous two sentences show that this is an evolving field.

The induction equation for MHD is written as

$$\frac{\partial \mathbf{B}}{\partial t} + \nabla \times \mathbf{E} = 0 \qquad (1)$$

with $\mathbf{E} = -\mathbf{v} \times \mathbf{B}$, where "$\mathbf{v}$" is the fluid velocity. In Fig. 1 we see that the components of the magnetic field "$\mathbf{B}$" (which are the primal variables of the scheme) are collocated at the faces of the mesh and the discrete divergence constraint is only preserved if the electric fields "$\mathbf{E}$" (which provide the update) are collocated at the edges of the mesh. But electromagnetics also tells us that the time rate of change of electromagnetic energy density is proportional to the divergence of the Poynting vector. In ideal MHD, the energy is concentrated in the magnetic field. The continuum version of Poynting's theorem for ideal MHD is given by

$$\partial_t \mathcal{E}_{mag} + \nabla \cdot \mathbf{S} = 0 \qquad \text{with} \qquad \mathcal{E}_{mag} \equiv \mathbf{B}^2/2 \qquad \text{and} \qquad \mathbf{S} \equiv \mathbf{E} \times \mathbf{B} \qquad (2)$$

Eqn. (2) is known as an involution in the mathematics literature. An involution is an inner property that the equations implicitly satisfy because of their intrinsic structure. In the next Section we will show that a certain discretization of the induction equation, i.e. eqn. (1), can be manipulated so that it has a discrete structure similar to eqn. (2). Such a numerical construct that allows additional properties of an involutive equation to be satisfied in a numerical code is known as a mimetic discretization. In other words, it makes a mime at the discrete, numerical level of the involutive property that is inherent in the partial differential equation. Unfortunately, up to now, no attempts have been made to mimetically incorporate the involution inherent in Poynting's Theorem into numerical MHD codes.

It is, however, likely that many practitioners have observed the following fact. A hydrodynamical code can stably simulate very high Mach number flows if the code is properly formulated. (Typically, hydrodynamical flows with Mach 100 are easy to simulate.) These are flows where the kinetic energy is ~10,000 times larger than the thermal energy. Most MHD codes falter when the magnetic energy becomes ~10,000 times larger than the thermal energy! Many practitioners might have been prompted to ask, "Why is there this asymmetry between kinetic and magnetic forms of energy?". We assert that it is because Poytning's theorem has not been consistently (mimetically) accounted for in numerical MHD and RMHD codes. The goal of this paper is to show that there is an easy way out. We assert that there is an inconsistency between the edge-centered electric fields in Fig. 1 and the face-centered electric field that is used for the update of the total energy density. This will be further detailed in the next Section where we will show that there is a simple way to restore this consistency. The nature of the correction is such that when the magnetic energy is somewhat comparable to the thermal energy, the correction does virtually nothing. Yet, when the magnetic energy is much larger than the thermal energy, small changes in the flux for total energy density can have a substantial influence on the positivity of the gas pressure. The suggested fix from this paper is very easy to apply as an add-on in all numerical MHD and RMHD codes. For the sake of completeness, it is worth mentioning that the nature of this inconsistency (though not the cure) was first mentioned to the first author by D. Spicer. Therefore, it is possible that some space physicists who used FCT methods might have known about this deficiency. The math as well as the development of the idea as it applies to modern higher order Godunov schemes is original to this short paper.

Because most astrophysical codes are based on a finite volume formulation, this paper pertains to finite volume schemes for MHD. Section II illustrates the nature of the problem and also provides a solution. Section III discusses higher order extensions of the insights generated in Setion II. Sectioon IV provides a step-by-step presentation of a physical constraint preserving (PCP) scheme. Section V shows several stringent test problems that work when this fix is incorporated. Section VI offers some quick conclusions.

## II) The Crux of the Problem – Consistency of Energy Flux with Poynting's Theorem

This is perhaps the most important Section of the paper. In Sub-section II.a we make an intuitive demonstration of the importance of mimetically incorporating consistency with Poynting's theorem in numerical MHD and RMHD codes. In Sub-section II.b we show how this is implemented in second order accurate MHD and RMHD codes for astrophysics and space physics. In Sub-section II.c we explain why this method works.

### II.a) Importance of Retaining Consistency with Poynting's Theorem

We provide this description within the context of MHD, but RMHD also benefits from the same insights. The energy flux for MHD can be written as $\left(e+\rho\mathbf{v}^2/2+P\right)\mathbf{v}+\mathbf{E}\times\mathbf{B}/\left(4\pi\right)$ with $\mathbf{E}=-\mathbf{v}\times\mathbf{B}$. Here "$e$" is the thermal energy density for MHD, "$\rho$" is the fluid density, "$P$" is the fluid pressure, "**v**" is the velocity vector, "**B**" is the magnetic field vector and "**E**" is the electric field vector. As a result, we realize that "$\mathbf{E} \times \mathbf{B} / (4\pi)$" is the Poynting flux. (The division of the square of the magnetic field by $4\pi$ is only needed in cgs units.) Notice, however, that the electric field vector is also given by the induction equation, i.e. eqn. (1), and is defined at the edges of the mesh with the help of the 2D Riemann solver (Balsara 2010, 2012, 2014, Balsara and Nkonga 2017, Balsara *et al*. 2025a, 2025b). The flux component for the evolution of the energy density is given by the 1D Riemann solver at the faces of the mesh. This is the crux of the problem. The face-centered electric fields that are used in the flux for magnetic energy density have no connection with the edge-centered electric fields that are used to update the facial components of the magnetic field.

We now show, via use of Fig. 1, that this consistency can be established. Consider the discrete update of the facial magnetic field components in Fig. 1. For a zone with sides $\Delta x$, $\Delta y$ and $\Delta z$ and time step $\Delta t$, the update from time $t^n$ to time $t^{n+1} = t^n + \Delta t$ can be written as:-

$$\frac{B_1^{x;n+1} - B_1^{x;n}}{\Delta t} = -\frac{1}{\Delta y}\left(E_4^z - E_1^z\right) + \frac{1}{\Delta z}\left(E_2^y - E_1^y\right) \tag{3}$$

$$\frac{B_2^{x;n+1} - B_2^{x;n}}{\Delta t} = -\frac{1}{\Delta y}\left(E_3^z - E_2^z\right) + \frac{1}{\Delta z}\left(E_3^y - E_4^y\right) \tag{4}$$

$$\frac{B_1^{y;n+1} - B_1^{y;n}}{\Delta t} = -\frac{1}{\Delta z}\left(E_4^x - E_1^x\right) + \frac{1}{\Delta x}\left(E_2^z - E_1^z\right) \tag{5}$$

$$\frac{B_2^{y;n+1} - B_2^{y;n}}{\Delta t} = -\frac{1}{\Delta z}\left(E_3^x - E_2^x\right) + \frac{1}{\Delta x}\left(E_3^z - E_4^z\right) \tag{6}$$

$$\frac{B_1^{z;n+1} - B_1^{z;n}}{\Delta t} = -\frac{1}{\Delta x}\left(E_4^y - E_1^y\right) + \frac{1}{\Delta y}\left(E_2^x - E_1^x\right) \tag{7}$$

$$\frac{B_2^{z;n+1} - B_2^{z;n}}{\Delta t} = -\frac{1}{\Delta x}\left(E_3^y - E_2^y\right) + \frac{1}{\Delta y}\left(E_3^x - E_4^x\right) \tag{8}$$

Our goal is to show that the above six equations implicitly ensure a discrete form of Poynting's theorem. We show this in the first order case. The zone-centered magnetic fields at the initial and final times of this timestep can be written as

$$\begin{aligned}
B_{avg}^{x;n+1} &\equiv \left(B_1^{x;n+1} + B_2^{x;n+1}\right)/2 \quad ; \quad B_{avg}^{x;n} \equiv \left(B_1^{x;n} + B_2^{x;n}\right)/2 \\
B_{avg}^{y;n+1} &\equiv \left(B_1^{y;n+1} + B_2^{y;n+1}\right)/2 \quad ; \quad B_{avg}^{y;n} \equiv \left(B_1^{y;n} + B_2^{y;n}\right)/2 \\
B_{avg}^{z;n+1} &\equiv \left(B_1^{z;n+1} + B_2^{z;n+1}\right)/2 \quad ; \quad B_{avg}^{z;n} \equiv \left(B_1^{z;n} + B_2^{z;n}\right)/2
\end{aligned} \tag{9}$$

As a result, the time rates of change for the zone-centered magnetic fields can be obtained by averaging the evolutionary equations for the face-centered magnetic fields. We get:-

$$\begin{aligned}
\frac{B_{avg}^{x;n+1} - B_{avg}^{x;n}}{\Delta t} &= -\frac{1}{2\Delta y}\left(E_4^z - E_1^z\right) + \frac{1}{2\Delta z}\left(E_2^y - E_1^y\right) - \frac{1}{2\Delta y}\left(E_3^z - E_2^z\right) + \frac{1}{2\Delta z}\left(E_3^y - E_4^y\right) \\
\frac{B_{avg}^{y;n+1} - B_{avg}^{y;n}}{\Delta t} &= -\frac{1}{2\Delta z}\left(E_4^x - E_1^x\right) + \frac{1}{2\Delta x}\left(E_2^z - E_1^z\right) - \frac{1}{2\Delta z}\left(E_3^x - E_2^x\right) + \frac{1}{2\Delta x}\left(E_3^z - E_4^z\right) \\
\frac{B_{avg}^{z;n+1} - B_{avg}^{z;n}}{\Delta t} &= -\frac{1}{2\Delta x}\left(E_4^y - E_1^y\right) + \frac{1}{2\Delta y}\left(E_2^x - E_1^x\right) - \frac{1}{2\Delta x}\left(E_3^y - E_2^y\right) + \frac{1}{2\Delta y}\left(E_3^x - E_4^x\right)
\end{aligned} \tag{10}$$

To get a measure of the change in magnetic energy, we multiply the first equation in eqn. (10) by $\left(B_{avg}^{x;n+1} + B_{avg}^{x;n}\right)/2$. Likewise, multiply the second equation in eqn. (10) by $\left(B_{avg}^{y;n+1} + B_{avg}^{y;n}\right)/2$ ; and multiply the third equation in eqn. (10) by $\left(B_{avg}^{z;n+1} + B_{avg}^{z;n}\right)/2$. We use the identity $\left(a+b\right)\left(a-b\right) = a^2 - b^2$ and add the resulting equations. The magnetic energy at the new and old times can be written as:-

$$\mathcal{E}_{mag}^{n+1} \equiv \left[\left(B_{avg}^{x;n+1}\right)^2 + \left(B_{avg}^{y;n+1}\right)^2 + \left(B_{avg}^{z;n+1}\right)^2\right]/2 \quad ; \quad \mathcal{E}_{mag}^{n} \equiv \left[\left(B_{avg}^{x;n}\right)^2 + \left(B_{avg}^{y;n}\right)^2 + \left(B_{avg}^{z;n}\right)^2\right]/2 \tag{11}$$

The time rate of change of the discrete magnetic energy at each zone-center can be written as:-

$$\frac{\mathcal{E}_{mag}^{n+1}-\mathcal{E}_{mag}^{n}}{\Delta t}=\frac{\left(B_{avg}^{x;n+1}+B_{avg}^{x;n}\right)}{2}\left[-\frac{1}{2\Delta y}\left(E_4^z-E_1^z\right)+\frac{1}{2\Delta z}\left(E_2^y-E_1^y\right)-\frac{1}{2\Delta y}\left(E_3^z-E_2^z\right)+\frac{1}{2\Delta z}\left(E_3^y-E_4^y\right)\right]$$
$$+\frac{\left(B_{avg}^{y;n+1}+B_{avg}^{y;n}\right)}{2}\left[-\frac{1}{2\Delta z}\left(E_4^x-E_1^x\right)+\frac{1}{2\Delta x}\left(E_2^z-E_1^z\right)-\frac{1}{2\Delta z}\left(E_3^x-E_2^x\right)+\frac{1}{2\Delta x}\left(E_3^z-E_4^z\right)\right]$$
$$+\frac{\left(B_{avg}^{z;n+1}+B_{avg}^{z;n}\right)}{2}\left[-\frac{1}{2\Delta x}\left(E_4^y-E_1^y\right)+\frac{1}{2\Delta y}\left(E_2^x-E_1^x\right)-\frac{1}{2\Delta x}\left(E_3^y-E_2^y\right)+\frac{1}{2\Delta y}\left(E_3^x-E_4^x\right)\right] \quad (12)$$

The above equation, by itself, does not provide any insight; it is just an intermediary step. However, it can be manipulated into a form that is highly suggestive of Poynting's Theorem as follows:-

$$\frac{\mathcal{E}_{mag}^{n+1}-\mathcal{E}_{mag}^{n}}{\Delta t}+\frac{1}{\Delta x}\left\{\begin{array}{l}\left[\frac{\left(E_3^y+E_4^y\right)}{2}\frac{\left(B_{avg}^{z;n+1}+B_{avg}^{z;n}\right)}{2}-\frac{\left(E_3^z+E_2^z\right)}{2}\frac{\left(B_{avg}^{y;n+1}+B_{avg}^{y;n}\right)}{2}\right]\\-\left[\frac{\left(E_2^y+E_1^y\right)}{2}\frac{\left(B_{avg}^{z;n+1}+B_{avg}^{z;n}\right)}{2}-\frac{\left(E_4^z+E_1^z\right)}{2}\frac{\left(B_{avg}^{y;n+1}+B_{avg}^{y;n}\right)}{2}\right]\end{array}\right\}$$
$$+\frac{1}{\Delta y}\left\{\begin{array}{l}\left[\frac{\left(E_3^z+E_4^z\right)}{2}\frac{\left(B_{avg}^{x;n+1}+B_{avg}^{x;n}\right)}{2}-\frac{\left(E_3^x+E_2^x\right)}{2}\frac{\left(B_{avg}^{z;n+1}+B_{avg}^{z;n}\right)}{2}\right]\\-\left[\frac{\left(E_2^z+E_1^z\right)}{2}\frac{\left(B_{avg}^{x;n+1}+B_{avg}^{x;n}\right)}{2}-\frac{\left(E_4^x+E_1^x\right)}{2}\frac{\left(B_{avg}^{z;n+1}+B_{avg}^{z;n}\right)}{2}\right]\end{array}\right\}$$
$$+\frac{1}{\Delta z}\left\{\begin{array}{l}\left[\frac{\left(E_3^x+E_4^x\right)}{2}\frac{\left(B_{avg}^{y;n+1}+B_{avg}^{y;n}\right)}{2}-\frac{\left(E_3^y+E_2^y\right)}{2}\frac{\left(B_{avg}^{x;n+1}+B_{avg}^{x;n}\right)}{2}\right]\\-\left[\frac{\left(E_2^x+E_1^x\right)}{2}\frac{\left(B_{avg}^{y;n+1}+B_{avg}^{y;n}\right)}{2}-\frac{\left(E_4^y+E_1^y\right)}{2}\frac{\left(B_{avg}^{x;n+1}+B_{avg}^{x;n}\right)}{2}\right]\end{array}\right\}=0 \quad (13)$$

Indeed, eqn. (13) is a discrete form of Poynting's theorem. It clearly identifies the discrete form of Poynting flux (i.e. $\mathbf{E}\times\mathbf{B}$) that needs to be used at face-centers to ensure that the zone-centered magnetic energy evolves consistently with the edge-based electric fields in Fig. 1. Notice that the edge-centered electric fields have to be used in eqn. (13) in order for it to be consistent with Poynting's theorem. Using face-centered electric fields will not give the desired consistency, This is the crux of the problem.

## II.b) Retaining Consistency with Poynting's Theorem in Second Order Schemes

Second order schemes still form the backbone for most astrophysical MHD and RMHD simulations. Therefore, in this Sub-section, we focus on how the consistency with Poynting's theorem should be implemented in second order schemes. In Section III we will discuss higher order extensions. In Section IV we will describe an end-to-end second order scheme that incorporates the ideas from this Sub-section.

Let us focus on Fig. 1 and introduce a slight enhancement of the notation. All the edge-centered electric fields will have a double starred superscript to denote that they have been obtained at the edges of the mesh using a two-dimensional Riemann solver. The y- and z-components of the edge-averaged electric fields at the center of the right x-face can be written with second order accuracy as

$$E_{fc}^{z**} = \frac{1}{2}\left( E_3^{z**} + E_2^{z**} \right) \quad ; \quad E_{fc}^{y**} = \frac{1}{2}\left( E_3^{y**} + E_4^{y**} \right) \tag{14}$$

Where the subscript "*fc*" denotes an edge-centered electric field that has been averaged to the face center. However, at the x-face-center we will also be invoking a one-dimensional Riemann solver which gives us a resolved state and a resolved flux. Let the outputs of the one-dimensional Riemann solver be denoted by a single starred superscript. From the resolved state we can extract the resolved magnetic fields at the face centers given by $B_{fc}^{y*}$ and $B_{fc}^{z*}$. From the resolved flux, we can use the dualism between electric fields and flux components (Balsara and Spicer 1999) to extract $E_{fc}^{y*}$ and $E_{fc}^{z*}$. The Poynting vector-based modification that one needs to add to the flux of total energy at the right x-face in Fig. 1 is now given by

$$\left[ \left( E_{fc}^{y**} - E_{fc}^{y*} \right) B_{fc}^{z*} - \left( E_{fc}^{z**} - E_{fc}^{z*} \right) B_{fc}^{y*} \right] \Big/ (4\pi) \tag{15}$$

(The division by $4\pi$ is only needed in cgs units and we may optionally dispense with it for the rest of the narrative.) This modification is second order accurate because eqn. (14) is also second order accurate. When the flow is smooth, and the magnetic fields are of modest strength, we know that $E_{fc}^{y**} \rightarrow E_{fc}^{y*}$ and $E_{fc}^{z**} \rightarrow E_{fc}^{z*}$ so that the change produced by eqn. (15) is negligible. That is why, most MHD codes work quite well most of the time when the magnetic energy is not too large compared to the thermal energy. It is only when the magnetic energy is very large compared to the thermal energy that eqn. (15) makes a substantial difference which nudges the solution towards a much-desired consistency with Poynting's Theorem!

We also realize that any good two-dimensional Riemann solver should also return resolved states for the transverse components of the magnetic field (see Balsara *et al*. 2025a,b). As a result, at each z-edge of Fig. 1, we will also have a resolved $B^{x**}$ and $B^{y**}$ in addition to having a resolved $E^{z**}$. Likewise, at each y-edge of Fig. 1, we will also have a resolved $B^{x**}$ and $B^{z**}$ in addition to having a resolved $E^{y**}$. In that case, we can come up with a sharper formulation for the correction term in eqn. (15) which we write as

$$\left[\frac{1}{2}\left(E_3^{y**}B_3^{z**}+E_4^{y**}B_4^{z**}\right)-\frac{1}{2}\left(E_3^{z**}B_3^{y**}+E_2^{z**}B_2^{y**}\right)\right]\Big/(4\pi)-\left[E_{fc}^{y*}B_{fc}^{z*}-E_{fc}^{z*}B_{fc}^{y*}\right]\Big/(4\pi) \qquad (16)$$

At second order of accuracy, eqns. (15) and (16) are both consistent with an approximation of the Poynting fluxes in eqn. (13). This completes our description of the second order accurate formulation.

**II.c) Why this Works?**

It is very desirable to generate insight as to why this mimetic enforcement of Poynting's theorem works. Let us first focus on non-relativistic MHD. The energy density is given by

$$\varepsilon=\frac{P}{\Gamma-1}+\frac{1}{2}\rho\mathbf{v}^2+\frac{\mathbf{B}^2}{8\pi} \qquad (17)$$

Here "$\varepsilon$" is the energy density, "$\rho$" is the density, "*P*" is the pressure, "**v**" is the velocity vector, "**B**" is the magnetic field vector and "$\Gamma$" is the ratio of specific heats. Now realize that the zone-centered magnetic field at second order is obtained from the averages of the facial magnetic field components, see Fig. 1. Eqn. (13) shows that as long as the Poynting flux at face-centers is fully consistent with the edge-centered electric fields, the zone-centered magnetic energy and its change is fully accounted for by Poynting's theorem. As a result, the rest of the numerical flux contributes only to the change in the thermal and kinetic energy densities. This is exactly what also happens in Euler flow. We know that for Euler flow, as long as the reconstructed second order variables in the flow code remain TVD and physical condition preserving, the pressure can be guaranteed to remain positive. The previous sentence is true when positivity preserving Riemann solvers, such as the HLL and LLF Riemann solvers, are used. As a result, we expect that for non-relativistic MHD the mimetic enforcement of Poynting's theorem keeps the numerical solution physical condition preserving without any need to degrade the timestep.

The explanation in the above paragraph also explains why these insights do not truly extend past second order. When we go past second order, the zone-centered magnetic field is not given by the averages of the facial magnetic field components. As a result, eqn. (13) ceases to strictly hold. Even so, we provide higher order extensions of these ideas in Section III.

Now let us focus on RMHD. In that case, the energy density can be analogously written as

$$\mathcal{E} = \rho h \gamma^2 - P + \frac{\mathbf{B}^2}{2} + \frac{\mathbf{v}^2\mathbf{B}^2 - \left(\mathbf{v}\cdot\mathbf{B}\right)^2}{2} \quad \text{with } \gamma \equiv \frac{1}{\sqrt{1-\mathbf{v}^2}} \quad \text{and } \rho h \equiv \rho + \frac{\Gamma}{\Gamma - 1} P \tag{18}$$

The speed of light is assumed unity, and the same notation as in eqn. (17) is used here. We can write the above equation as

$$\mathcal{E} = \rho \gamma^2 + \left[\frac{\Gamma\left(\gamma^2 - 1\right) + 1}{\Gamma - 1}\right] P + \frac{\mathbf{B}^2}{2}\left[1 + \mathbf{v}^2\left(1 - \cos^2\theta\right)\right] \quad \text{with} \quad \cos^2\theta \equiv \frac{\left(\mathbf{v}\cdot\mathbf{B}\right)^2}{\left(\mathbf{v}^2\mathbf{B}^2\right)} \tag{19}$$

We now see that eqn. (13) does not apply exactly in the RMHD case. This happens especially in situations where $\mathbf{v}^2\left(1-\cos^2\theta\right)$ is comparable to unity, which happens when the magnetic field and velocity are increasingly orthogonal to one another. In practice, at second order, we have still found that the second order RMHD formulation described in Section IV still retains its physical condition preserving property with no compromise in the CFL number. But the discussion in this paragraph also enables us to understand why higher order methods for RMHD, which allow substantial variation in $\cos^2\theta$ within a zone, may in some rare instances need the timestep to be lowered. In Sub-section IV.b we discuss a method that can preempt the need for timestep reduction.

**III) Higher Order Extensions**

Eqns. (15) and (16) have given us the crucial insight. We are using edge-centered electric and magnetic fields from a two-dimensional Riemann solver to update the face-centered magnetic fields. These are then used to obtain the zone-centered magnetic fields that are used for extracting the gas pressure from the energy density. However, in the past, it was traditional to use only the one-dimensional Riemann solvers to update the total energy density. As a result, the update was inconsistent with Poynting's theorem. Our new insight from eqns. (14) through (16) is that we can use the edge-centered electric fields from the two-dimensional Riemann solver to obtain a face-centered Poynting flux. This Poynting flux can be used to replace the Poynting flux from the one-dimensional Riemann solver. This modification only changes the component of the flux that is used to update the energy density. These extensions can be carried out at higher order, as shown in this Section, but the extensions have to be done delicately so that the higher order accuracy is not damaged by the modification.

Fig. 2 shows us the right x-face that is associated with Fig. 1. In this figure, we show the electric fields not just in that x-face, but we also show relevant electric fields in the four x-faces that are adjacent to the face in question. The electric fields obtained from the edges of those faces will also contribute to the (better than second order accurate) Poynting flux that we will build using just the results from the two-dimensional Riemann solvers. We also show the nine nodal points within the x-face where one-dimensional Riemann solvers are invoked in the x-direction at third and fourth order of accuracy. The information from these one-dimensional Riemann solvers can also be used to build a Poynting flux. Realize, therefore, that we will have two Poynting fluxes at each zone boundary – one that is based on two-dimensional Riemann solvers at the edges of the mesh and another one that is based on one-dimensional Riemann solvers at the nodes within the faces of the mesh. Our eventual objective is to subtract off the latter Poynting flux and replace it with the former Poynting flux.

Fig. 3 shows us the same x-face as in Fig. 2, but now we are looking down on the x-axis as we view that x-face. The facial nodes have been numbered in order to make it easier to write explicit expressions for the Poynting flux. Any term that relies on a one-dimensional Riemann solver is shown with a single starred superscript. Any term that relies on a two-dimensional Riemann solver is shown with a double starred superscript. We also remind the reader that any one-dimensional Riemann solver for MHD will return the resolved state, including magnetic field components, as well as the resolved flux, which includes the transverse electric fields. We also remind the reader that any good two-dimensional Riemann solver that is invoked at the edges of the mesh should return not just the electric field along the edge but also the two magnetic field components that are transverse to that edge. For example, see the two-dimensional Riemann solvers designed in Balsara *et al*. (2025a,b).

Let us first build an area-integrated Poynting flux in the x-direction that is based on the one-dimensional Riemann solver, because that is simpler to understand. The nodal points shown in Fig. 3 correspond to Simpson's rule, as a result, we use the corresponding weights to get

$$\begin{aligned} S^{x}_{1D-RS} = &\frac{1}{36}\left(E^{y*}_1 B^{z*}_1 - E^{z*}_1 B^{y*}_1\right) + \frac{4}{36}\left(E^{y*}_2 B^{z*}_2 - E^{z*}_2 B^{y*}_2\right) + \frac{1}{36}\left(E^{y*}_3 B^{z*}_3 - E^{z*}_3 B^{y*}_3\right) \\ &+ \frac{4}{36}\left(E^{y*}_4 B^{z*}_4 - E^{z*}_4 B^{y*}_4\right) + \frac{16}{36}\left(E^{y*}_5 B^{z*}_5 - E^{z*}_5 B^{y*}_5\right) + \frac{4}{36}\left(E^{y*}_6 B^{z*}_6 - E^{z*}_6 B^{y*}_6\right) \\ &+ \frac{1}{36}\left(E^{y*}_7 B^{z*}_7 - E^{z*}_7 B^{y*}_7\right) + \frac{4}{36}\left(E^{y*}_8 B^{z*}_8 - E^{z*}_8 B^{y*}_8\right) + \frac{1}{36}\left(E^{y*}_9 B^{z*}_9 - E^{z*}_9 B^{y*}_9\right) \end{aligned} \quad (20)$$

The $S^{x}_{1D-RS}$ denotes that it is the x-directional component of the Poynting flux obtained from the one-dimensional Riemann solver. It is the well-known fourth order accurate quadrature formula for two-dimensional integration based on a tensor product Simpson rule. The numerical subscripts for "*E*" and "*B*" in eqn. (20) indicate the nodal points in Fig. 3 where they are obtained from the one-dimensional Riemann solvers.

Now let us turn our attention to the more intricate task of building the Poynting flux at the x-face using just the electric field that is longitudinal to each edge and the two magnetic field components that are transverse to each edge. In this paragraph we focus on fourth order accuracy. Since these are the outputs from the two-dimensional Riemann solver, in Fig. 3 we show that along each y-edge we can construct the line integral of the product of $E^{y**}$ and $B^{z**}$. Similarly, along each z-edge we can construct the line integral of the product of $E^{z**}$ and $B^{y**}$. These integrals are shown by the angled brackets in Fig. 3. Therefore, our first instinct would be to write the x-component of the Poynting flux by simply borrowing ideas from eqn. (16) as

$$\frac{1}{2}\left(\left\langle E_3^{y**} B_3^{z**} \right\rangle + \left\langle E_4^{y**} B_4^{z**} \right\rangle\right) - \frac{1}{2}\left(\left\langle E_3^{z**} B_3^{y**} \right\rangle + \left\langle E_2^{z**} B_2^{y**} \right\rangle\right) \tag{21}$$

The above result, while nominally consistent, is not accurate enough. The reason is that the above expression is only second order accurate while eqn. (20) is up to fourth order accurate. As a result, even for very smooth flow, they would not cancel in an order of accuracy preserving fashion. To be genuinely fourth order accurate, eqn. (21) needs to take on higher order derivative terms. However, when the flow is truly non-smooth, those higher order derivative terms can generate spurious oscillations which should be suppressed. To balance the conflicting needs of higher order accuracy and oscillation-free evolution, we can define a flattener function $\phi$ at the face in question. Also see Balsara (2012b) for the description of a flattener function. The flattener function is zero when the solution is smooth and can become one when the solution is non-smooth. We then get a higher order accurate (i.e. fourth order accurate) representation of the Poynting flux, which we write as

$$S^x_{2D-RS} = \left\langle E^{y**} B^{z**} \right\rangle_{x-face-average} - \left\langle E^{z**} B^{y**} \right\rangle_{x-face-average}$$

with

$$\begin{aligned} \left\langle E^{y**} B^{z**} \right\rangle_{x-face-average} &\equiv \frac{1}{2}\left(\left\langle E_3^{y**} B_3^{z**} \right\rangle + \left\langle E_4^{y**} B_4^{z**} \right\rangle\right) \\ &- \frac{1}{24}(1-\phi)\left(\left\langle E_5^{y**} B_5^{z**} \right\rangle - \left\langle E_3^{y**} B_3^{z**} \right\rangle - \left\langle E_4^{y**} B_4^{z**} \right\rangle + \left\langle E_6^{y**} B_6^{z**} \right\rangle\right) \end{aligned} \tag{22}$$

and

$$\begin{aligned} \left\langle E^{z**} B^{y**} \right\rangle_{x-face-average} &\equiv \frac{1}{2}\left(\left\langle E_3^{z**} B_3^{y**} \right\rangle + \left\langle E_2^{z**} B_2^{y**} \right\rangle\right) \\ &- \frac{1}{24}(1-\phi)\left(\left\langle E_6^{z**} B_6^{y**} \right\rangle - \left\langle E_3^{z**} B_3^{y**} \right\rangle - \left\langle E_2^{z**} B_2^{y**} \right\rangle + \left\langle E_5^{z**} B_5^{y**} \right\rangle\right) \end{aligned}$$

Here, $S^x_{2D-RS}$ denotes that it is the x-directional component of the Poynting flux that is obtained from the two-dimensional Riemann solver. (The extra terms in eqn. (22), as compared to eqn. (21), are second derivatives of the edge-integrated quantities. They raise the accuracy in the vicinity of smooth flow from second order accuracy to fourth order accuracy.) The final step is to add $\left(S^x_{2D-RS} - S^x_{1D-RS}\right)$ to the x-component of the flux for energy density. Similar terms can be added to the y-and z-components of the fluxes for energy density.

It is also interesting to show that the ideas in the above paragraph are extensible to even higher orders; i.e. that they are generalizable. Therefore, in Fig. 4, we show the extension to sixth order. We do not explicitly show the extension of eqn. (20) to sixth order since that is a well-known two-dimensional quadrature formula involving tensor product Gauss-Lobatto nodes. The extension of eqn. (22) to sixth order, however, requires some non-trivial innovation. The final result, after a little searching, is

$$S^x_{2D-RS} = \left\langle E^{y**} B^{z**} \right\rangle_{x-face-average} - \left\langle E^{z**} B^{y**} \right\rangle_{x-face-average}$$

with

$$\begin{aligned}
\left\langle E^{y**} B^{z**} \right\rangle_{x-face-average} &\equiv \frac{1}{2}\left( \left\langle E_3^{y**} B_3^{z**} \right\rangle + \left\langle E_4^{y**} B_4^{z**} \right\rangle \right) \\
&- \frac{1}{24}(1-\phi)\left( \left\langle E_5^{y**} B_5^{z**} \right\rangle - \left\langle E_3^{y**} B_3^{z**} \right\rangle - \left\langle E_4^{y**} B_4^{z**} \right\rangle + \left\langle E_6^{y**} B_6^{z**} \right\rangle \right) \\
&- \frac{11}{1440}(1-\phi)\left( -\left\langle E_7^{y**} B_7^{z**} \right\rangle + 3\left\langle E_5^{y**} B_5^{z**} \right\rangle - 2\left\langle E_3^{y**} B_3^{z**} \right\rangle - 2\left\langle E_4^{y**} B_4^{z**} \right\rangle + 3\left\langle E_6^{y**} B_6^{z**} \right\rangle - \left\langle E_8^{y**} B_8^{z**} \right\rangle \right)
\end{aligned}$$

and

$$\begin{aligned}
\left\langle E^{z**} B^{y**} \right\rangle_{x-face-average} &\equiv \frac{1}{2}\left( \left\langle E_3^{z**} B_3^{y**} \right\rangle + \left\langle E_2^{z**} B_2^{y**} \right\rangle \right) \\
&- \frac{1}{24}(1-\phi)\left( \left\langle E_6^{z**} B_6^{y**} \right\rangle - \left\langle E_3^{z**} B_3^{y**} \right\rangle - \left\langle E_2^{z**} B_2^{y**} \right\rangle + \left\langle E_5^{z**} B_5^{y**} \right\rangle \right) \\
&- \frac{11}{1440}(1-\phi)\left( -\left\langle E_8^{z**} B_8^{y**} \right\rangle + 3\left\langle E_6^{z**} B_6^{y**} \right\rangle - 2\left\langle E_3^{z**} B_3^{y**} \right\rangle - 2\left\langle E_2^{z**} B_2^{y**} \right\rangle + 3\left\langle E_5^{z**} B_5^{y**} \right\rangle - \left\langle E_7^{z**} B_7^{y**} \right\rangle \right)
\end{aligned} \tag{23}$$

Eqn. (23) shows that a systematic process exists for extending these ideas to all orders of accuracy. While we have only focused on applications that go up to fourth order of accuracy in this paper, the math in this paragraph shows that there are no restrictions to our Poynting flux correction approach at even higher orders.

This completes our description of the higher order extension for restoring consistency with Poynting's theorem in astrophysical MHD and RMHD.

## IV) PCP Methods with Poynting Flux Consistency and 2D Riemann Solver

Methods that preserve the physical realizability of a PDE system, such as positive densities and pressures, are known as physical condition preserving (PCP). Notice that eqn. (13) is formally second order accurate for two reasons. First, the Poynting fluxes are centered in time, which means that it is possible to use single step update strategies like ADER (Dumbser *et al*. 2008, 2013, Balsara *et al*. 2009, 2013). We have also used Strong Stability Preserving Runge-Kutta (SSP-RK) methods and found that they perform equally well. Second, realize that the zone-centered magnetic field is the arithmetic average of the facial magnetic fields only at first or second order, and eqn. (13) relies on that insight in order to remain valid. As a result, we expect our proposed Poynting flux correction to strictly help with the PCP property only at first and second order. Indeed, we have verified that all first and second order schemes that we have formulated with the Poynting flux correction remain PCP. For that reason Sub-section IV.a describes a second order PCP method, whereas Sub-section IV.b describes higher order extensions.

**IV.a) Second order PCP Method with Poynting Flux Consistency and 2D Riemann Solver**

We present here a very Simple PCP-preserving, Poynting-flux-consistent, single step, second order scheme that uses one-dimensional and two-dimensional Riemann solvers. The scheme we describe is of predictor corrector type and incorporates Poynting's Theorem, as well as PCP properties, in each of the predictor and corrector steps. We describe the process of going from time $t^n$ to time $t^{n+1} = t^n + \Delta t$ on a uniform 3D mesh with zone sizes $\Delta x$, $\Delta y$ and $\Delta z$. We will use integer subscripts to indicate mesh centers and half integer subscripts to indicate half zone displacements from zone centers. We generically label zone centers as $(i, j, k)$. We start with zone-centered conserved variables $\bar{\mathbf{U}}^n_{i,j,k}$ and facial variables $\bar{B}^{x;n}_{i+1/2,j,k}$, $\bar{B}^{y;n}_{i,j+1/2,k}$, $\bar{B}^{z;n}_{i,j,k+1/2}$ at the face-centers of the mesh. We take the zone-centered conserved variables and the face-centered magnetic field components as our primal variables. The overbars indicate averages over the volume of the zones or zone-faces, as appropriate. We also assume that the last three components of $\bar{\mathbf{U}}^n_{i,j,k}$ have been loaded with the arithmetic averages of the facial magnetic fields, as is common for a second order scheme. The construction is described in the following steps:-

**Step 1)** From the conserved variables $\bar{\mathbf{U}}_{i,j,k}^{n}$ obtain the zone-centered primitive variables $\mathbf{V}_{i,j,k}^{n}$. At second order, this is trivial. At each zone center obtain a flattener, see Balsara (2012), to identify whether the zone is overrun by a shock or very close to one.

**Step 2)** Obtain the undivided differences $\Delta_x \mathbf{V}_{i,j,k}^{n}$, $\Delta_y \mathbf{V}_{i,j,k}^{n}$, $\Delta_z \mathbf{V}_{i,j,k}^{n}$ by applying a limiter in each direction. In each face, obtain the undivided differences of the facial magnetic field component in the two transverse directions. For example, for $\bar{B}_{i+1/2,j,k}^{x;n}$, obtain $\Delta_y B_{i+1/2,j,k}^{x;n}$ and $\Delta_z B_{i+1/2,j,k}^{x;n}$. Use the flattener to flatten these undivided differences if shocks are present. The limiter can be any TVD limiter. (However, if a higher quality reconstruction is needed, one can also use the linear part of the r=3 WENO scheme from Section 2.1 of Balsara, Garain and Shu (2016) if one wants a better quality interpolation. But note the caveat that these higher quality reconstructions may not have the best PCP properties.)

**Step 3)** Now make a PCP requirement on the above undivided differences. At a minimum, make sure that the interpolated values at face-centers, i.e. $\mathbf{V}_{i,j,k}^{n} \pm \Delta_x \mathbf{V}_{i,j,k}^{n} / 2$, $\mathbf{V}_{i,j,k}^{n} \pm \Delta_y \mathbf{V}_{i,j,k}^{n} / 2$ and $\mathbf{V}_{i,j,k}^{n} \pm \Delta_z \mathbf{V}_{i,j,k}^{n} / 2$, have positive densities and pressures; if not, reduce the undivided differences to make it so. This PCP procedure can be easily added to any code. (The PCP requirement can even be applied to edge-centers and vertex-centers if more stabilization is desired; for example, see the Barth limiter in eqns. (3.41) to (3.47) of Balsara (2004). For the problems presented here we did not need this additional stabilization from the Barth limiter.)

**Step 4)** At each zone center, use $\mathbf{V}_{i,j,k}^{n}$ to evaluate zone-centered electric fields $E_{i,j,k}^{x;n}$, $E_{i,j,k}^{y;n}$ and $E_{i,j,k}^{z;n}$. From these electric fields, build their undivided differences by applying a TVD limiter to these zone-centered electric fields. For example, for $E_{i,j,k}^{x;n}$ we build $\Delta_x E_{i,j,k}^{x;n}$, $\Delta_y E_{i,j,k}^{x;n}$ and $\Delta_z E_{i,j,k}^{x;n}$. Also apply the flattener to these undivided differences.

**Step 5)** We are now ready to make the first call to the 1D Riemann solvers for the predictor step which we denote $RS_{1D}$. We obtain face-centered fluxes at time $t^n$ given by

$$
\begin{aligned}
&\mathbf{F}^{*;n}_{i+1/2,j,k} = RS_{1D}\left(\mathbf{V}^{n}_{i,j,k} + \Delta_x \mathbf{V}^{n}_{i,j,k}/2; \mathbf{V}^{n}_{i+1,j,k} - \Delta_x \mathbf{V}^{n}_{i+1,j,k}/2\right) \quad ; \\
&\mathbf{G}^{*;n}_{i,j+1/2,k} = RS_{1D}\left(\mathbf{V}^{n}_{i,j,k} + \Delta_y \mathbf{V}^{n}_{i,j,k}/2; \mathbf{V}^{n}_{i,j+1,k} - \Delta_y \mathbf{V}^{n}_{i,j+1,k}/2\right) \quad ; \\
&\mathbf{H}^{*;n}_{i,j,k+1/2} = RS_{1D}\left(\mathbf{V}^{n}_{i,j,k} + \Delta_z \mathbf{V}^{n}_{i,j,k}/2; \mathbf{V}^{n}_{i,j,k+1} - \Delta_z \mathbf{V}^{n}_{i,j,k+1}/2\right)
\end{aligned}
\tag{24}
$$

The inputs to the one-dimensional Riemann solvers in the above equation are second order accurate in space. We should also save the face-centered resolved states from these one-dimensional Riemann solvers, i.e. $\mathbf{U}^{*;n}_{i+1/2,j,k}$ , $\mathbf{U}^{*;n}_{i,j+1/2,k}$ and $\mathbf{U}^{*;n}_{i,j,k+1/2}$ , which will be needed for the Poynting flux correction in eqns. (15) or (16). It is also important to make sure that the one-dimensional Riemann solver retains the continuity of the longitudinal component of the magnetic field at whichever face it is invoked. HLL or LLF Riemann solvers can be used because they guarantee a PCP flux. (However, if a higher quality 1D Riemann solver that is complete in the 1D wave structure is desired, please try the one from Dumbser and Balsara (2016). But note the caveat that only the numerical fluxes from HLL and LLF Riemann solvers are PCP.)

**Step 6)** We now describe the process of feeding the multidimensional Riemann solver that is invoked at the edges of the mesh for the predictor step. This is done here by instantiating the process of obtaining the resolved longitudinal electric field $E^{z;**;n}_{i+1/2,j+1/2,k}$ as well as the resolved transverse magnetic fields, $B^{x;**;n}_{i+1/2,j+1/2,k}$ and $B^{y;**;n}_{i+1/2,j+1/2,k}$ , at the z-edge. Please also see the input variables that go into a two-dimensional Riemann solver as shown in Fig. 5 (which is copied here from Fig. 2 of Balsara *et al*. (2025a) for the sake of completeness). We first interpolate the four zone-centered z-components of the electric field to the edge in question as

$$
\begin{aligned}
&E_{zLD} = E^{z;n}_{i,j,k} + \Delta_x E^{z;n}_{i,j,k}/2 + \Delta_y E^{z;n}_{i,j,k}/2 \quad ; \quad E_{zRD} = E^{z;n}_{i+1,j,k} - \Delta_x E^{z;n}_{i+1,j,k}/2 + \Delta_y E^{z;n}_{i+1,j,k}/2 \quad ; \\
&E_{zLU} = E^{z;n}_{i,j+1,k} + \Delta_x E^{z;n}_{i,j+1,k}/2 - \Delta_y E^{z;n}_{i,j+1,k}/2 \quad ; \quad E_{zRU} = E^{z;n}_{i+1,j+1,k} - \Delta_x E^{z;n}_{i+1,j+1,k}/2 - \Delta_y E^{z;n}_{i+1,j+1,k}/2
\end{aligned}
\tag{25}
$$

We then interpolate the four face-centered components of the magnetic field to the same edge as

$$
\begin{aligned}
&B_{xD} = \overline{B}^{x;n}_{i+1/2,j,k} + \Delta_y B^{x;n}_{i+1/2,j,k}/2 \quad ; \quad B_{xU} = \overline{B}^{x;n}_{i+1/2,j+1,k} - \Delta_y B^{x;n}_{i+1/2,j+1,k}/2 \quad ; \\
&B_{yL} = \overline{B}^{y;n}_{i,j+1/2,k} + \Delta_x B^{y;n}_{i,j+1/2,k}/2 \quad ; \quad B_{yR} = \overline{B}^{y;n}_{i+1,j+1/2,k} - \Delta_x B^{y;n}_{i+1,j+1/2,k}/2 \quad ;
\end{aligned}
\tag{26}
$$

The interpolations described in eqns. (25) and (26) are second order accurate in space. Once the above electric and magnetic field components are in hand, they serve as inputs to the two-dimensional Riemann solver. For a two-dimensional HLL Riemann solver see eqns. (3.6), (3.7), (3.8) and (3.10) from Balsara *et al*. (2025a). For a two-dimensional Riemann solver that is complete in all the wave modes in both directions, please see Balsara and Nkonga (2017). For the simplest LLF Riemann solver in two-dimensions, please see eqn. (3.9) from Balsara *et al*. (2025a) which we repeat here for the sake of convenience:-

$$
\begin{aligned}
B^{x;**;n}_{i+1/2,j+1/2,k} &= \left(B_{xU}+B_{xD}\right)/2+\left(E_{zLD}-E_{zLU}+E_{zRD}-E_{zRU}\right)/\left(4S\right) \quad ; \\
B^{y;**;n}_{i+1/2,j+1/2,k} &= \left(B_{yR}+B_{yL}\right)/2+\left(E_{zRD}-E_{zLD}+E_{zRU}-E_{zLU}\right)/\left(4S\right) \quad ; \\
E^{z;**;n}_{i+1/2,j+1/2,k} &= \left(E_{zRU}+E_{zLU}+E_{zLD}+E_{zRD}\right)/4+S\left[B_{xD}-B_{xU}+B_{yR}-B_{yL}\right]/2 \quad .
\end{aligned}
\tag{27}
$$

Here, "$S$" is the maximal speed from the four 1D Riemann solvers that abut the edge in question. The resolved edge-centered variables from eqn. (27) (and other formulae like it) should be saved because we now use them to apply Poynting vector corrections to the numerical fluxes $\mathbf{F}^{*;n}_{i+1/2,j,k}$, $\mathbf{G}^{*;n}_{i,j+1/2,k}$ and $\mathbf{H}^{*;n}_{i,j,k+1/2}$ that were obtained in the previous step; see eqns. (15) or (16). Just like the PCP interpolation, this is a very small but extraordinarily essential procedure that is needed in a good MHD scheme. (We are literally adding one well-designed number to each of the three fluxes at each of the three face-centers!)

**Step 7)** Now that we have the Poynting vector corrected fluxes in hand, we can build the zone-centered undivided difference of the conserved variables in time as follows

$$
\Delta_t \mathbf{U}^{n}_{i,j,k} = -\frac{\Delta t}{\Delta x}\left(\mathbf{F}^{*;n}_{i+1/2,j,k}-\mathbf{F}^{*;n}_{i-1/2,j,k}\right)-\frac{\Delta t}{\Delta y}\left(\mathbf{G}^{*;n}_{i,j+1/2,k}-\mathbf{G}^{*;n}_{i,j-1/2,k}\right)-\frac{\Delta t}{\Delta z}\left(\mathbf{H}^{*;n}_{i,j,k+1/2}-\mathbf{H}^{*;n}_{i,j,k-1/2}\right) \tag{28}
$$

This gives a prediction for the time rate of change of the zone-centered conserved variables that is spatially second order accurate. From the above, we can build a predicted conserved variable at time $t^{n+1}$ as $\tilde{\mathbf{U}}^{n+1}_{i,j,k} \equiv \bar{\mathbf{U}}^{n}_{i,j,k}+\Delta_t \mathbf{U}^{n}_{i,j,k}$. From $\tilde{\mathbf{U}}^{n+1}_{i,j,k}$ we can easily build the vector of predicted primitive variables as $\tilde{\mathbf{V}}^{n+1}_{i,j,k}$. From that, we can build the undivided difference of the primitive variables in time as $\Delta_t \mathbf{V}^{n}_{i,j,k} \equiv \tilde{\mathbf{V}}^{n+1}_{i,j,k}-\mathbf{V}^{n}_{i,j,k}$ .

**Step 8)** At each of the faces of the mesh, we can resort to equations that resemble eqns. (3) to (8) to find the time rate of change of the facial magnetic field. For example, for the x-component $\bar{B}^{x;n}_{i+1/2,j,k}$ we obtain

$$\Delta_t B^{x;n}_{i+1/2,j,k} = -\frac{\Delta t}{\Delta y}\left(E^{z;**;n}_{i+1/2,j+1/2,k} - E^{z;**;n}_{i+1/2,j-1/2,k}\right) + \frac{\Delta t}{\Delta z}\left(E^{y;**;n}_{i+1/2,j,k+1/2} - E^{y;**;n}_{i+1/2,j,k-1/2}\right) \tag{29}$$

This gives a prediction for the time rate of change of the facial magnetic field components that is spatially second order accurate. Notice that it is obtained exclusively from the edge-centered electric fields, which play an important role for Poynting flux consistency.

**Step 9)** Now that we have $\tilde{\mathbf{V}}^{n+1}_{i,j,k}$ and $\mathbf{V}^{n}_{i,j,k}$ we can build zone-centered electric fields at both those times. This allows us to obtain a finite difference-based prediction for the time rate of change of the zone-centered electric fields; i.e. $\Delta_t E^{x;n}_{i,j,k}$, $\Delta_t E^{y;n}_{i,j,k}$ and $\Delta_t E^{z;n}_{i,j,k}$. This completes the predictor step. Realize, therefore, that by the end of the predictor step, we have already incorporated PCP properties into the reconstruction as well as Poynting flux consistency for the numerical flux.

**Step 10)** Our next steps will consist of making a time centered corrector step so that we have second order consistency, as well as consistency with eqn. (13). To retain consistency with eqn. (13) we will, of course, have to retain Poynting flux consistency for the numerical flux even in the corrector step. To that end, we make another invocation of the one-dimensional Riemann solvers at the faces of the mesh, but this time we do it in a time-centered fashion. We have

$$\begin{aligned}
\mathbf{F}^{*;n+1/2}_{i+1/2,j,k} &= RS_{1D}\left(\mathbf{V}^{n}_{i,j,k} + \Delta_x \mathbf{V}^{n}_{i,j,k}/2 + \Delta_t \mathbf{V}^{n}_{i,j,k}/2 ; \mathbf{V}^{n}_{i+1,j,k} - \Delta_x \mathbf{V}^{n}_{i+1,j,k}/2 + \Delta_t \mathbf{V}^{n}_{i+1,j,k}/2\right) \;;\\
\mathbf{G}^{*;n+1/2}_{i,j+1/2,k} &= RS_{1D}\left(\mathbf{V}^{n}_{i,j,k} + \Delta_y \mathbf{V}^{n}_{i,j,k}/2 + \Delta_t \mathbf{V}^{n}_{i,j,k}/2 ; \mathbf{V}^{n}_{i,j+1,k} - \Delta_y \mathbf{V}^{n}_{i,j+1,k}/2 + \Delta_t \mathbf{V}^{n}_{i,j+1,k}/2\right) \;;\\
\mathbf{H}^{*;n+1/2}_{i,j,k+1/2} &= RS_{1D}\left(\mathbf{V}^{n}_{i,j,k} + \Delta_z \mathbf{V}^{n}_{i,j,k}/2 + \Delta_t \mathbf{V}^{n}_{i,j,k}/2 ; \mathbf{V}^{n}_{i,j,k+1} - \Delta_z \mathbf{V}^{n}_{i,j,k+1}/2 + \Delta_t \mathbf{V}^{n}_{i,j,k+1}/2\right)
\end{aligned} \tag{30}$$

In contrast to eqn. (24), the above formulae hold true at time $t^{n+1/2} = t^n + \Delta t/2$ because the input states to the one-dimensional Riemann solver are time-centered as well as being second order accurate in space and time. As in the predictor step, so in the corrector step, we should also save the face-centered resolved states, i.e. $\mathbf{U}^{*;n+1/2}_{i+1/2,j,k}$ , $\mathbf{U}^{*;n+1/2}_{i,j+1/2,k}$ and $\mathbf{U}^{*;n+1/2}_{i,j,k+1/2}$ , which will be needed for the Poynting flux correction in eqns. (15) or (16). As before, use HLL or LLF Riemann solvers, though better options exist. The longitudinal component of the magnetic field should also be kept continuous in the Riemann solver.

**Step 11)** We now describe the process of feeding the multidimensional solver for the corrector step by instantiating the process of obtaining the resolved longitudinal electric field $E^{z;**;n+1/2}_{i+1/2,j+1/2,k}$ as well as the resolved transverse magnetic fields, $B^{x;**;n+1/2}_{i+1/2,j+1/2,k}$ and $B^{y;**;n+1/2}_{i+1/2,j+1/2,k}$ , at the z-edge. Analogous to eqns. (25) and (26), we define time-centered inputs for the two-dimensional Riemann solver as

$$\begin{aligned}
E_{zLD} &= E^{z;n}_{i,j,k} + \Delta_x E^{z;n}_{i,j,k}/2 + \Delta_y E^{z;n}_{i,j,k}/2 + \Delta_t E^{z;n}_{i,j,k}/2 \quad ; \\
E_{zRD} &= E^{z;n}_{i+1,j,k} - \Delta_x E^{z;n}_{i+1,j,k}/2 + \Delta_y E^{z;n}_{i+1,j,k}/2 + \Delta_t E^{z;n}_{i+1,j,k}/2 \quad ; \\
E_{zLU} &= E^{z;n}_{i,j+1,k} + \Delta_x E^{z;n}_{i,j+1,k}/2 - \Delta_y E^{z;n}_{i,j+1,k}/2 + \Delta_t E^{z;n}_{i,j+1,k}/2 \quad ; \\
E_{zRU} &= E^{z;n}_{i+1,j+1,k} - \Delta_x E^{z;n}_{i+1,j+1,k}/2 - \Delta_y E^{z;n}_{i+1,j+1,k}/2 + \Delta_t E^{z;n}_{i+1,j+1,k}/2
\end{aligned} \tag{31}$$

and

$$\begin{aligned}
B_{xD} &= \bar{B}^{x;n}_{i+1/2,j,k} + \Delta_y B^{x;n}_{i+1/2,j,k}/2 + \Delta_t B^{x;n}_{i+1/2,j,k}/2 \quad ; \quad B_{xU} = \bar{B}^{x;n}_{i+1/2,j+1,k} - \Delta_y B^{x;n}_{i+1/2,j+1,k}/2 + \Delta_t B^{x;n}_{i+1/2,j+1,k}/2 \quad ; \\
B_{yL} &= \bar{B}^{y;n}_{i,j+1/2,k} + \Delta_x B^{y;n}_{i,j+1/2,k}/2 + \Delta_t B^{y;n}_{i,j+1/2,k}/2 \quad ; \quad B_{yR} = \bar{B}^{y;n}_{i+1,j+1/2,k} - \Delta_x B^{y;n}_{i+1,j+1/2,k}/2 + \Delta_t B^{y;n}_{i+1,j+1/2,k}/2 \quad ;
\end{aligned} \tag{32}$$

The interpolations described in eqns. (31) and (32) are second order accurate in space and time. Learning from eqn. (27), if we put the time-centered eqns. (31) and (32) into the formula for a multidimensional Riemann solver, we get time-centered electric fields at the edges of the mesh. The analogue of eqn. (27) is

$$\begin{aligned}
B^{x;**;n+1/2}_{i+1/2,j+1/2,k} &= \left(B_{xU} + B_{xD}\right)/2 + \left(E_{zLD} - E_{zLU} + E_{zRD} - E_{zRU}\right)/\left(4S\right) \quad ; \\
B^{y;**;n+1/2}_{i+1/2,j+1/2,k} &= \left(B_{yR} + B_{yL}\right)/2 + \left(E_{zRD} - E_{zLD} + E_{zRU} - E_{zLU}\right)/\left(4S\right) \quad ; \\
E^{z;**;n+1/2}_{i+1/2,j+1/2,k} &= \left(E_{zRU} + E_{zLU} + E_{zLD} + E_{zRD}\right)/4 + S\left[B_{xD} - B_{xU} + B_{yR} - B_{yL}\right]/2 \quad .
\end{aligned} \tag{33}$$

The variable "*S*" is again the extremal speed from the abutting 1D Riemann solvers from eqn. (30).The resolved edge-centered variables from eqn. (33) (and other formulae like it) should be saved because we now use them to apply Poynting vector corrections to the numerical fluxes $\mathbf{F}^{*;n+1/2}_{i+1/2,j,k}$ , $\mathbf{G}^{*;n+1/2}_{i,j+1/2,k}$ and $\mathbf{H}^{*;n+1/2}_{i,j,k+1/2}$ that were obtained in the previous step; see eqns. (15) or (16). As before, this is a very small but extraordinarily essential procedure that is needed in a good MHD scheme. (We are literally adding one well-designed number to each of the three fluxes at each of the three face-centers!)

**Step 12)** We are now ready to institute the final part of the corrector step update. The zone-centered variables can take a time-centered second order accurate in space and time update as follows

$$\bar{\mathbf{U}}_{i,j,k}^{n+1} = \bar{\mathbf{U}}_{i,j,k}^{n} - \frac{\Delta t}{\Delta x}\left(\mathbf{F}_{i+1/2,j,k}^{*;n+1/2} - \mathbf{F}_{i-1/2,j,k}^{*;n+1/2}\right) - \frac{\Delta t}{\Delta y}\left(\mathbf{G}_{i,j+1/2,k}^{*;n+1/2} - \mathbf{G}_{i,j-1/2,k}^{*;n+1/2}\right) - \frac{\Delta t}{\Delta z}\left(\mathbf{H}_{i,j,k+1/2}^{*;n+1/2} - \mathbf{H}_{i,j,k-1/2}^{*;n+1/2}\right) \quad (34)$$

As before, we instantiate the time-centered second order accurate in space and time update for the x-component of the face-centered field as

$$\bar{B}_{i+1/2,j,k}^{x;n+1} = \bar{B}_{i+1/2,j,k}^{x;n} - \frac{\Delta t}{\Delta y}\left(E_{i+1/2,j+1/2,k}^{z;**;n+1/2} - E_{i+1/2,j-1/2,k}^{z;**;n+1/2}\right) + \frac{\Delta t}{\Delta z}\left(E_{i+1/2,j,k+1/2}^{y;**;n+1/2} - E_{i+1/2,j,k-1/2}^{y;**;n+1/2}\right) \quad (35)$$

This completes our description of the second order space-time accurate scheme which is PCP and consistent with Poynting's Theorem. We have been able to run the most stringent of 2D test problems with a CFL of 0.4, as shown in the next Sub-section. For 3D tests we recommend reducing the CFL to 0.25. The method works for MHD and RMHD as shown in the ensuing test problems. The second order PCP methods for MHD and RMHD described in this Sub-section can, in and of themselves, be very interesting to many practitioners. They require minimal changes to pre-existing higher order Godunov solvers; yet their effect is quite significant!

One variation on the method described in this Sub-section is worth mentioning. In eqns. (30), (31) and (32) we re-used the spatial gradients that were built at time $t^n$. This was done for the sake of economy. However, note that we can also build $\mathbf{U}_{i,j,k}^{n+1/2} = \mathbf{U}_{i,j,k}^{n} + \Delta_t \mathbf{U}_{i,j,k}^{n} / 2$ and facial terms like $B_{i+1/2,j,k}^{x;n+1/2} = B_{i+1/2,j,k}^{x;n} + \Delta_t B_{i+1/2,j,k}^{x;n} / 2$. The facial magnetic field components can again be averaged to the zone-centered magnetic fields, so that the last three magnetic components of $\mathbf{U}_{i,j,k}^{n+1/2}$ are overwritten. We can then use $\mathbf{U}_{i,j,k}^{n+1/2}$ to obtain time-centered primitive variables $\mathbf{V}_{i,j,k}^{n+1/2}$. These primitive variables can then be used to build undivided differences using a TVD reconstruction. The time-centered reconstructed variables can now be used to form time-centered fluxes and electric fields. The Poynting fix is also applied again. This is a slightly more expensive option because it involves two TVD reconstruction steps.

### IV.b) Strategies for Making Higher Order Finite Volume Codes PCP

It is worth recalling that a second order scheme produces results that are dramatically better than a first order scheme. The difference between a first order scheme and a higher order scheme is starkly, and jarringly, apparent. If sufficient resolution is used, the difference between a second order scheme and a third or fourth order scheme is indeed visible, but quite small. Third and fourth order schemes can also benefit from the Poynting flux correction, but they cannot be made to guarantee the PCP property (see Sub-section II.c). This is because eqn. (13), as shown, cannot be used to claim the PCP property beyond second order. However, in order to ensure PCP evolution at all orders, we can do the next best thing:- In Balsara *et al*. (2025a,b) we showed that a higher order scheme can be hybridized with a first order scheme in order to make it PCP. Here we suggest that since the second order scheme for MHD and RMHD that is described in the previous Sub-section is so lightweight and PCP, it can also be hybridized with a higher order scheme in order to guarantee the PCP property. Because the difference between a second order scheme and a higher order scheme is quite modest, it seamlessly improves the quality of the higher order method. We also point out that for problems that are not very extreme, the hybridization is completely unobtrusive and it is such that the full advantage of the higher order scheme is retained. The methods presented here are also single-step ADER-WENO methods with the result that they minimize the messaging on parallel supercomputers.

We now describe in detail a strategy that improves on the one described in Balsara *et al*. (2025a,b). Fig. 6a shows a tent function in two dimensions. It is unity in the zone of interest, shown with a grey dot. The other vertices that define the tent function are shown with blue dots in Fig. 6a along with their coordinates, which are shown in red. The tent function $\phi(x,y)$ allows us to apply a taper of two zones on either side of a two-dimensional mesh in the xy-plane, as shown in Fig. 6a. (One can envision a broader tent function if one wishes; sometimes we found it useful to have a tent function with three zones.) It is now easy to envision a three-dimensional tent function that is analogous to Fig. 6a. We will illustrate the idea in two dimensions because it is easier. The core idea is that we decide ahead of time that the astrophysical or space physics code works saliently when the pressure "$P$" is more than a factor $\chi_{hi}$ of the total energy density "$\varepsilon$". We also establish that the code runs into problems when the pressure "$P$" is less than a factor $\chi_{lo}$ of the total energy density "$\varepsilon$". Typically, one can take $\chi_{hi} = 0.05$ and $\chi_{lo} = 0.001$. Also, in each zone $(i, j)$ we build the ratio $\chi_{i,j} \equiv P_{i,j} / \varepsilon_{i,j}$ . As a result, to each zone that has a low pressure we can ascribe the tent function with a variable height given by

$$\tilde{\phi}_{i,j}(x,y) \equiv \min\left(1.0, \max\left(0.0, \frac{(\chi_{hi} - \chi_{i,j})}{(\chi_{hi} - \chi_{lo})}\right)\right)\phi\left(x - x_i, y - y_j\right) \tag{36}$$

Notice that $\tilde{\phi}_{i,j}(x,y)$ is a tent function with solution-dependent height centered at a problematic zone with center at $(x_i, y_j)$. Now, within each zone $(i, j)$ we construct a value $\tilde{\tilde{\phi}}_{i,j}$ which is the maximum of all the neighboring tent functions.

Now realize that the higher order reconstruction is indeed the problem. Therefore, the final reconstruction within a zone can be taken as $\tilde{\tilde{\phi}}_{i,j}$ times the second order reconstruction plus $\left(1 - \tilde{\tilde{\phi}}_{i,j}\right)$ times the higher order reconstruction. When a zone is problem-free, the fully higher order reconstruction is retained. For zones that are problematic (i.e. with uncomfortably low pressures), one reverts to second order accuracy. This ensures that the reconstruction within a zone remains salient. The same thing is done for the reconstruction of the magnetic field components at the faces. If needed, one can also use methods from Balsara (2012) to ensure a physical reconstruction within each zone. Note though that ensuring a PCP reconstruction in a numerical scheme that is better than second order accurate does not *per se* ensure a PCP update. It just ensures that two states that go into a Riemann solver from either side of a zone boundary are PCP. As a result, a PCP preserving Riemann solver (like HLL or LLF) will return physical output fluxes and electric fields. (The Poynting fix is of course applied to the energy flux.) But the update that one gets from such fluxes and electric fields is still not guaranteed to keep the updated solution PCP!

That is where we need the support from the second order PCP scheme described in Sub-section IV.a. If a zone has an update that violates the PCP condition then we iteratively give that zone a greater fraction of the second order fluxes and electric fields and a smaller fraction of the higher order fluxes and electric fields. (The two fractions should add up to unity.) This can also be done iteratively and in a progressive fashion; but for that we have found it useful to use the two-dimensional top hat function shown in Fig. 6b. This top hat function is also easily extended to three dimensions. Since the second order scheme is PCP, it ensures that at some point in the iteration, the result will be PCP in all zones.

The previous three paragraphs provide, as it were, fixes of last resort if a higher order scheme is in the process of producing negative pressures. But there is a strategy that we describe here that tries to preempt that as much as possible. It is not 100% guaranteed to work, but it improves the solution substantially when it does. To understand why it works, let us understand what is going on in a higher order method. The three panels in Fig. 7 show four adjoining zones from the same mesh function. The dashed lines show the mesh function. The thick solid lines in Fig. 7a show the reconstructed function when a first order scheme is used. Notice that the jumps in the reconstructed function are very large, especially at the zone boundary of interest which is shown by an arrow. Fig. 7b shows the reconstructed function when a second order scheme is used. We see that the piecewise linear profiles have significantly reduced the jump at the zone boundary of interest. Fig. 7c shows the piecewise reconstructed function when a third order scheme (or a higher order scheme) is used. We see that the reconstructed profiles within the zones have reduced the jump at the zone boundary with the arrow even further. Since the dissipation provided by the Riemann solver is proportional to the jump in the reconstructed profile, we see that as we go to higher orders, the dissipation is reduced. The way a Riemann solver operates is such that a reduction in dissipation corresponds to a reduction in entropy generation at the zone boundary of interest. In a generic situation, reducing the entropy generation is a good thing; that is indeed why higher order schemes produce results that are superior to first and second order schemes in general. However, if the mesh function already has a zone with pressure that is close to zero, then reducing the entropy generation at the zone boundary can drive the pressure in that zone to negative values. We realize, therefore, that within zones that have a dangerously low pressure, the use of a high order scheme will inexorably drive the pressure to negative values in one or a few timesteps. We have to design a strategy that preempts that.

If the only method that we have at our disposal is a higher order scheme, then there are no good options left. However, we also have the support of the PCP-preserving second order scheme from Sub-section IV.a. So what we want to prevent is the use of higher order fluxes and electric fields in zones that simultaneously have too small a pressure and are also showing too little entropy generation. But we have the support of the second order scheme. Therefore, from each zone of the time-updated second order scheme, we can extract the second order entropy $S_{2nd;i,j}^{n+1} = P_{2nd;i,j}^{n+1} \Big/ \left( \rho_{2nd;i,j}^{n+1} \right)^{\Gamma}$ . We can also extract the entropy from the same zone that has been tentatively updated with a higher order scheme, i.e. $S_{hi;i,j}^{n+1} = P_{hi;i,j}^{n+1} \Big/ \left( \rho_{hi;i,j}^{n+1} \right)^{\Gamma}$ . We can also evaluate $\chi_{hi;i,j}^{n+1} \equiv P_{hi;i,j}^{n+1} \big/ \varepsilon_{hi;i,j}^{n+1}$ . We can then design a switch given by

$$\min\left( 1.0, \max\left( 0.0, \frac{\left( \chi_{hi} - \chi_{hi;i,j}^{n+1} \right)}{\left( \chi_{hi} - \chi_{lo} \right)} \right) \right) \left\{ \min\left[ 1.0, \max\left[ 0.0, \left( \left( S_{2nd;i,j}^{n+1} - S_{hi;i,j}^{n+1} \right) \big/ S_{2nd;i,j}^{n+1} - \Sigma_{\min} \right) \big/ \Delta\Sigma \right] \right] \right\} \tag{37}$$

Good choices include $\left( \Sigma_{\min}, \Delta\Sigma \right) = \left( 0.1, 0.2 \right)$ or $\left( \Sigma_{\min}, \Delta\Sigma \right) = \left( 0.05, 0.1 \right)$; in very intense situations one may even use $\left( \Sigma_{\min}, \Delta\Sigma \right) = \left( 0.02, 0.04 \right)$. The first two preempt the development of dangerously low pressures within a zone, the third does so even more aggressively. We have used the former two choices for most problems, but we used the third choice for the relativistic blast problem shown here. Notice that the above switch has two factors. The first factor, just like eqn. (36), only activates when the pressure is becoming dangerously small within a zone. The second factor only activates when the entropy of the higher order scheme is *decreasing* much faster than the entropy of the second order scheme, which is PCP. In most zones which have a robust pressure, this switch remains zero. Even if the entropy should increase in a zone with very small pressure, the switch remains zero. It only becomes greater than zero in situations where the pressure is dangerously low and the entropy is simultaneously decreasing too fast relative to the second order scheme, which is PCP. When that happens, we blend in second order fluxes and electric fields in that zone by a factor that is proportional to the value of eqn. (37). This is the last, physics-driven cure for zones that would produce a negative pressure in one or a few timesteps if only the higher order fluxes and electric fields had been used. This step is designed to use the second order PCP scheme as a guide to *preempt* the development of dangerously low or negative pressures in a higher order scheme.

## V) Results

### V.a) Accuracy Demonstration for MHD and RMHD

We first show the accuracy of our method at second, third and fourth order using the non-relativistic magnetized vortex (Balsara 2004) and relativistic magnetized vortex (Balsara and Kim 2016). For complete setup details of the vortex problems, we refer the reader to the above-mentioned references. Fig 8a shows the $L^1$-error plots for the non-relativistic magnetized vortex, and Fig 8b shows the $L^1$-error plots for the relativistic magnetized vortex. Fig 8c shows the $L^\infty$-error plots for the non-relativistic magnetized vortex, and Fig 8d shows the $L^\infty$-error plots for the relativistic magnetized vortex. These errors are shown for the $B_y$ variable (*y*-component of the magnetic field). In both cases, the expected convergence rate is recovered, showing that the proposed Poynting-flux correction does not damage the order property. For all the problems considered in this Sub-section, we have used a CFL number of 0.4.

### V.b) Nonrelativistic MHD in 2D

We next examine the robustness of the method using stringent nonrelativistic MHD test problems. These include an extremely strongly magnetized blast wave and an exceptionally high-Mach astrophysical jet with a very low plasma beta. All problems in this Sub-section were run with a CFL of 0.4.

For the two-dimensional MHD blast-wave test, we adopt the strongly magnetized configuration of Wu and Shu (2018), which is considerably more severe than the moderately magnetized cases considered by Balsara and Spicer (1999a, 1999b). The test is characterized by a very small plasma beta $\beta = 2.51\times10^{-6}$ and is solved on the domain $[-0.6, 0.6]\times[-0.6, 0.6]$. Initially, $\rho = 1$ and $\mathbf{v} = 0$ throughout the domain. The pressure is $p = 0.1$ everywhere except inside the central region with radius $r < 0.1$, where it is raised to an exceptionally high pressure of $p = 10^4$. A uniform magnetic field of strength $B_x = 1000$ is imposed in the $x$-direction. We run the simulation using the first, second and fourth-order schemes on a $400 \times 400$ mesh until $t = 0.001$. Figs. 9a, 9b and 9c show the logarithm (base 10) of the density profiles at first, second and fourth orders, and Figs. 9d, 9e and 9f show the logarithm (base 10) of the pressure profiles at first, second and fourth orders, respectively. We see that the resulting solutions are consistent with those reported by Wu and Shu (2018). It is worth emphasizing that visualizing the logarithm (base 10) of the density and pressure variables is a much more powerful demonstration of the success of the method because the curatives that are provided in problematic zones can leave their imprint on the solution. Those imprints, which usually show up at low values of density and pressure, come to the fore when log scaling is used. We see from Fig. 9 that the regions with low values of density and pressure also show clean and smooth profiles in the figure. To the best of our knowledge, this is the first paper to show the logarithms of the density and pressure variables.

We then consider the Mach 800 astrophysical jet of Balsara (2012b), using the magnetized formulation adopted by Wu and Shu (2018). Only the strongest magnetic case is examined, with $B_a = \sqrt{20000}$ where the corresponding plasma beta is $\beta_a = 10^{-4}$. The computational domain is $[-0.5, 0.5] \times [0, 1.5]$, while the remaining initial and boundary conditions follow Wu and Shu (2018). We run the simulation using the first, second and fourth-order schemes on a $400 \times 600$ mesh until $t = 0.002$. Figs. 10a, 10b and 10c show the logarithm (base 10) of the density profiles at the first, second and fourth orders, and Figs. 10d, 10e and 10f show the logarithm (base 10) of the pressure profiles at the first, second and fourth order, respectively. Even under this extreme magnetization, the jet's working surface and surrounding cocoon are sharply resolved at fourth order, demonstrating the robustness of the methods presented here.

**V.c) Relativistic MHD in 2D**

We now demonstrate the PCP capabilities of our methods for an ultra-relativistic high-Mach jet for the RMHD system. We were also able to demonstrate the successful operation of the method for a super-strong RMHD blast problem. We have used a CFL number of 0.4.

We first consider the RMHD Blast problem from Wu and Shu (2020). The initial pressure to magnetic pressure ratio for this problem is $\beta := 2p / B^2 = 2.5\times10^{-6}$. The problem is defined on the two-dimensional domain that spans $[-6,6]\times[-6,6]$. For $r < 0.8$ (where $r = \sqrt{x^2 + y^2}$), the density and pressure are $10^{-2}$ and 1, respectively. For $r > 1$, they are $10^{-4}$ and $5\times10^{-4}$. Between $r = 0.8$ and $r = 1$, both quantities decrease linearly. The magnetic field is directed along the *x*-axis with magnitude 20. The polytropic index is $\Gamma = 4/3$. We run the simulation using the first, second and fourth-order schemes on a $400\times400$ mesh until $t = 4$. Figs. 11a, 11b and 11c show the logarithm (base 10) of the density profiles at first, second and fourth orders, and Figs. 11d, 11e and 11f show the logarithm (base 10) of the pressure profiles at first, second and fourth orders, respectively. The numerical results are consistent with those reported in Wu and Shu (2020).

Now, we consider the relativistic astrophysical jet introduced by Wu and Shu (2020). The problem is defined on $[-12,12]\times[0,25]$ with a uniform ambient state given by $\rho = 1$, $\mathbf{v} = 0$, and $p = 2.3536241\times10^{-5}$. The magnetic field is initialized as $\mathbf{B} = (0, \sqrt{2000p}, 0)$. This configuration gives an initial Lorentz factor of approximately $\gamma \simeq 7.09$ and a relativistic Mach number $M_r \simeq 354.37$. At the lower boundary $y = 0$, the jet is injected through $|x| < 0.5$ with the state $(\rho, v_x, v_y, v_z, p, B_x, B_y, B_z) = (0.1, 0, 0.99, 0, 2.3536241\times10^{-5}, 0, \sqrt{2000p}, 0)$. Outflow conditions are imposed elsewhere, with $\gamma = 5/3$ (adiabatic constant) and the initial pressure to magnetic pressure ratio for this problem is $\beta := 2p / B^2 = 10^{-3}$. We run the simulation using the first, second and fourth-order schemes on a $480\times500$ mesh until $t = 0.002$. Figs. 12a, 12b and 12c show the logarithm (base 10) of the density profiles at the first, second and fourth orders, and Figs. 12d, 12e and 12f show the logarithm (base 10) of the pressure profiles at the first, second and fourth order, respectively. At fourth order, interfaces remain well resolved, and the overall solution are consistent with the results in Wu and Shu (2020).

### V.d) MHD and RMHD in 3D

Finally, to demonstrate the generality of the methods presented here, we consider the three-dimensional version of the two-dimensional MHD and RMHD Blast problems presented above. Both problems were run with a CFL of 0.25 in three-dimensions.

We first show the MHD Blast problem. As this is just a three-dimensional extension of the two-dimensional MHD problem, the pressure ratio between the initial blast pressure and the ambient pressure is taken given as 10,000:0.1. The initial plasma beta in the ambient medium for this problem is $\beta = 2.51 \times 10^{-6}$. We run the simulation using the third-order scheme on a $200 \times 200 \times 200$ zone mesh until $t = 0.001$. The obtained logarithm (base 10) of density and pressure profiles (clipped boxplots) are shown in Fig. 13. We see that the clipped plots (Fig. 13) are analogous to the two-dimensional plots presented in Fig. 9, showing the applicability of the schemes presented here in three dimensions.

Next, we consider the three-dimensional RMHD Blast problem, which is an extension of the two-dimensional RMHD Blast problem. The initial pressure to magnetic pressure ratio for this problem is $\beta := 2p / B^2 = 2.5 \times 10^{-6}$. We run this simulation using the third-order scheme on a $200 \times 200 \times 200$ zone mesh until a time of $t$=4.0. The obtained logarithm (base 10) of density and pressure profiles (clipped boxplots) are shown in Fig. 14. We see that the clipped plots (Fig. 14) are analogous to the two-dimensional plots presented in Fig. 11, showing the applicability of the schemes presented here for the three-dimensional RMHD equations.

## VI) Conclusions

In this paper we have shown that Poynting's theorem has involutive properties that can be mimetically incorporated in numerical MHD codes. This is accomplished by starting with the discrete versions of the induction equation, as shown in eqns. (3) to (8). Through a sequence of steps, we show in eqn. (13) that we obtain a discrete equation for the evolution of the zone-centered magnetic energy that uses a discrete divergence of a face-centered vector that very closely resembles Poynting's vector. This enables us to identify the discrete version of the Poynting flux. We see from eqns. (15) and (16) that the Poynting flux that should be used is based on the output of a two-dimensional Riemann solver that is invoked at the edges of the mesh. We realize that the numerical flux for the energy density that is obtained from a one-dimensional Riemann solver is indeed the wrong thing to use if we want mimetic consistency with Poynting's theorem. Fortunately, at second order the correction from the edge-centered electric and magnetic fields is easy and completely intuitive as shown in eqns. (15) and (16). At higher orders, eqns. (20) and (22) show us that a precise higher order extension of this correction can be found. The correction that we propose is extremely simple to implement and computationally inexpensive. We have applied this correction to several test problems that are used for accuracy analysis. We have shown that the accuracy is unaffected by this correction. Moreover, several stringent test problems that would have given current generations of MHD and RMHD codes difficulty now go through when this correction is incorporated. At least at second order, our experience has been that the process works perfectly. A very simple and inexpensive second order scheme is described in Sub-section IV.a which incorporates this Poynting flux correction and is, therefore, PCP. In Balsara *et al.* (2025a) we had also documented methods that ensure that a first order PCP method can be seamlessly hybridized with a higher order method in order to retrieve the PCP property in higher order methods. These hybridization methods have been extended in Sub-section IV.b of this paper so that a second order PCP method can be seamlessly hybridized with a higher order method. Since second order methods are closer in solution quality to higher order methods, the hybridization of the two methods is much smoother. All the methods presented here are also single-step finite volume methods with rather small stencils, with the result that they minimize the messaging. This makes them very suitable for parallel supercomputers.

Second order methods are still the workhorses in computational astrophysics and space physics. It is, therefore, possible that a practitioner might want just a very good and stable second order scheme. Within the context of second order schemes that retain PCP properties and consistency with Poynting's Theorem, we would recommend the use of TVD limiters with HLL or LLF fluxes for very stringent problems. But we realize that milder applications may not need strong PCP enforcement, though all applications would benefit from retaining consistency with Poynting's Theorem. In that case, it is feasible to swap out the TVD limiter in a second order scheme and replace it by the linear part of the r=3 WENO limiter from Balsara, Garain and Shu (2016). It is also possible to swap out HLL or LLF Riemann solvers and use instead a complete Riemann solver like the HLLI Riemann solver from Dumbser and Balsara (2016). As a result, within one and the same second order code base, by the use of eqns. (36) and (37) and the associated discussion, it is possible to have a single self-adjusting second order code which is very robust when a stringent application warrants it, and also very accurate when the application is not so stringent.

In this paper, we have focused on finite volume schemes for MHD and RMHD because they seem to be the most popular methods in the astrophysical literature. However, finite difference methods may also be of interest because of their very substantial speed advantage over finite volume methods. Progress on finite difference methods for MHD and RMHD, that incorporate multidimensional Riemann solvers and consistency with Poynting's theorem, will be reported in Balsara *et al*. (2026c). In a future paper we will also focus on introducing the Poynting flux correction to discontinuous Galerkin schemes for MHD. In this paper we have described the Poynting flux correction idea for ideal MHD, but the same idea should also apply to non-ideal MHD. Similarly, PIC codes lavish a lot of effort on energy-conserving particle integrators. But they do not pay any attention to total energy conservation in the plasma, which is made of particles and electromagnetic fields. It may be hoped that methods that retain consistency with Poynting's theorem for electrodynamics will also play a role in PIC codes.

In recent years, a lot of effort has gone into devising methods that try to preserve positivity of pressure in MHD simulations (Wu 2018, Wu and Shu 2018, 2019, 2023a,b, 2026, Yan *et al*. 2026). Many of those methods try to reduce the order of accuracy, and the proofs of the PCP property, such as they exist, are only restricted to first order schemes. Furthermore, as reported by Wei *et al*. (2026), the Geometric Quasi-Linearization method in Wu and Shu (2023b) is not locally flux conservative and "can occasionally produce incorrect jumps across strong shock discontinuities in some numerical simulations". The method presented here is very general and applies to MHD schemes of all orders of accuracy. Our method is also locally flux conservative at all orders of accuracy. We have even shown that it extends seamlessly to RMHD applications, thereby showing that it generalizes to other PDE systems with an induction equation. It is, therefore, possible that the PCP approaches mentioned at the beginning of this paragraph were missing an essential ingredient. We anticipate that the mathematical theorems that have been proved (for retaining the PCP property) would be vastly simplified if the Poynting flux correction-related insights in this paper are applied to them. The above-mentioned approaches also increase the signal speed in the Riemann solver while simultaneously restricting the timestep to a sixth of the allowable timestep in order to acquire their PCP capabilities. (Artificially increasing the speed in the Riemann solver also increases the dissipation.) In our approach, we have not seen the need for such timestep reduction or artificial increases in the wave speed.

This paper closes a conceptual gap that has long existed in numerical MHD and RMHD, and documents the substantial gains in doing so. The most important significance of this work to astrophysics and space physics is, of course, that problems with strong magnetic fields that were out of scope have now come within scope. This work might also be useful in the simulation of plasma confinement problems and plasma propulsion problems.

**Acknowledgements**

DSB acknowledges support via NSF grants NSF-SNSF award AST-2434532 and NSF_OAC-2514038. DB acknowledges support via NFSG, Grant N6/25/1079.

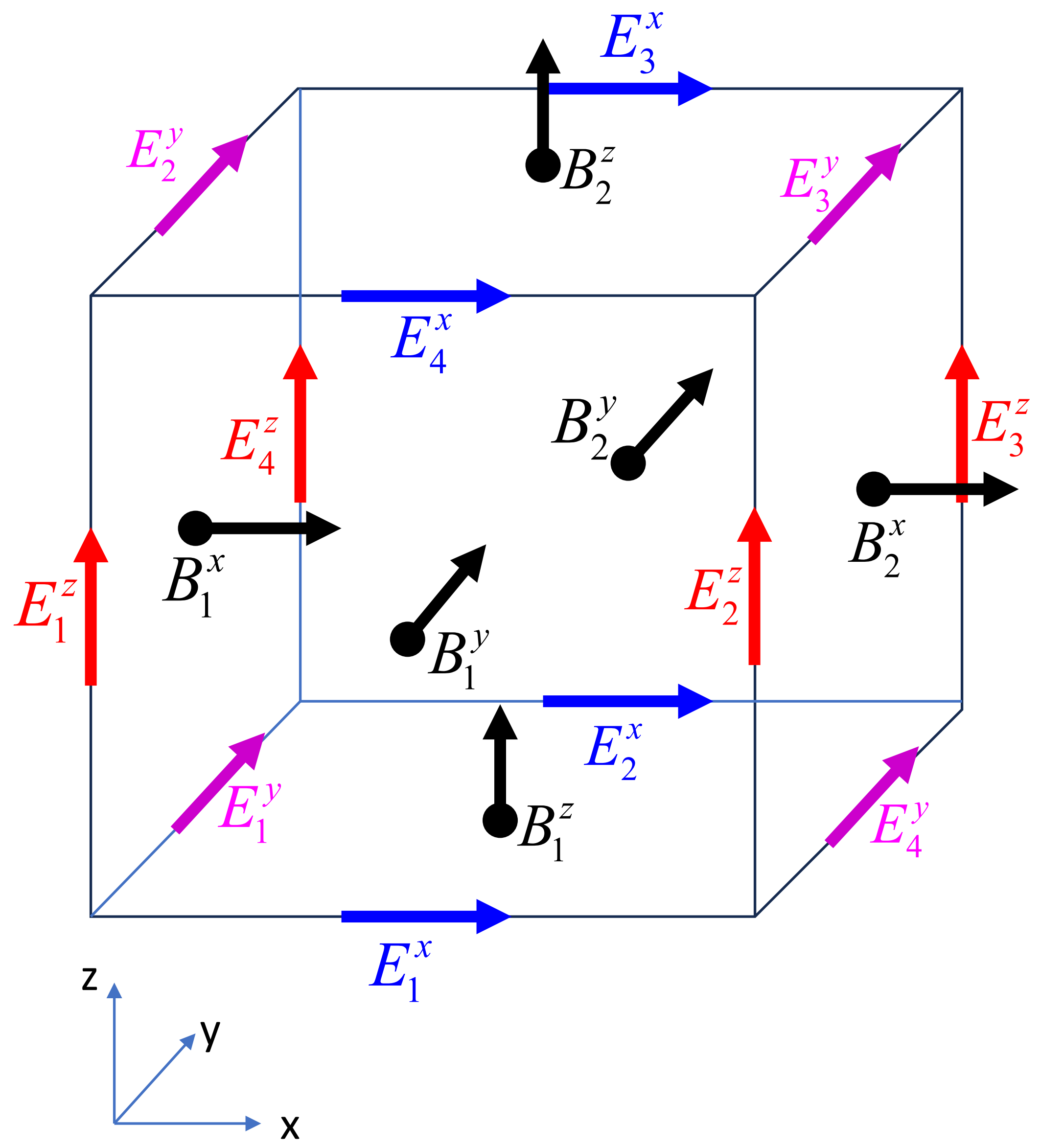

*Fig. 1 shows the arrangement of face-centered magnetic fields and edge-centered electric fields in a layout and numbering scheme that is especially useful for showing the discrete form of Poynting's theorem for magnetic energy conservation. The facial magnetic field components are shown in black. The z-components of the electric field are shown in red; the y-components of the electric field are shown in magenta; the x-components of the electric field are shown in blue.*

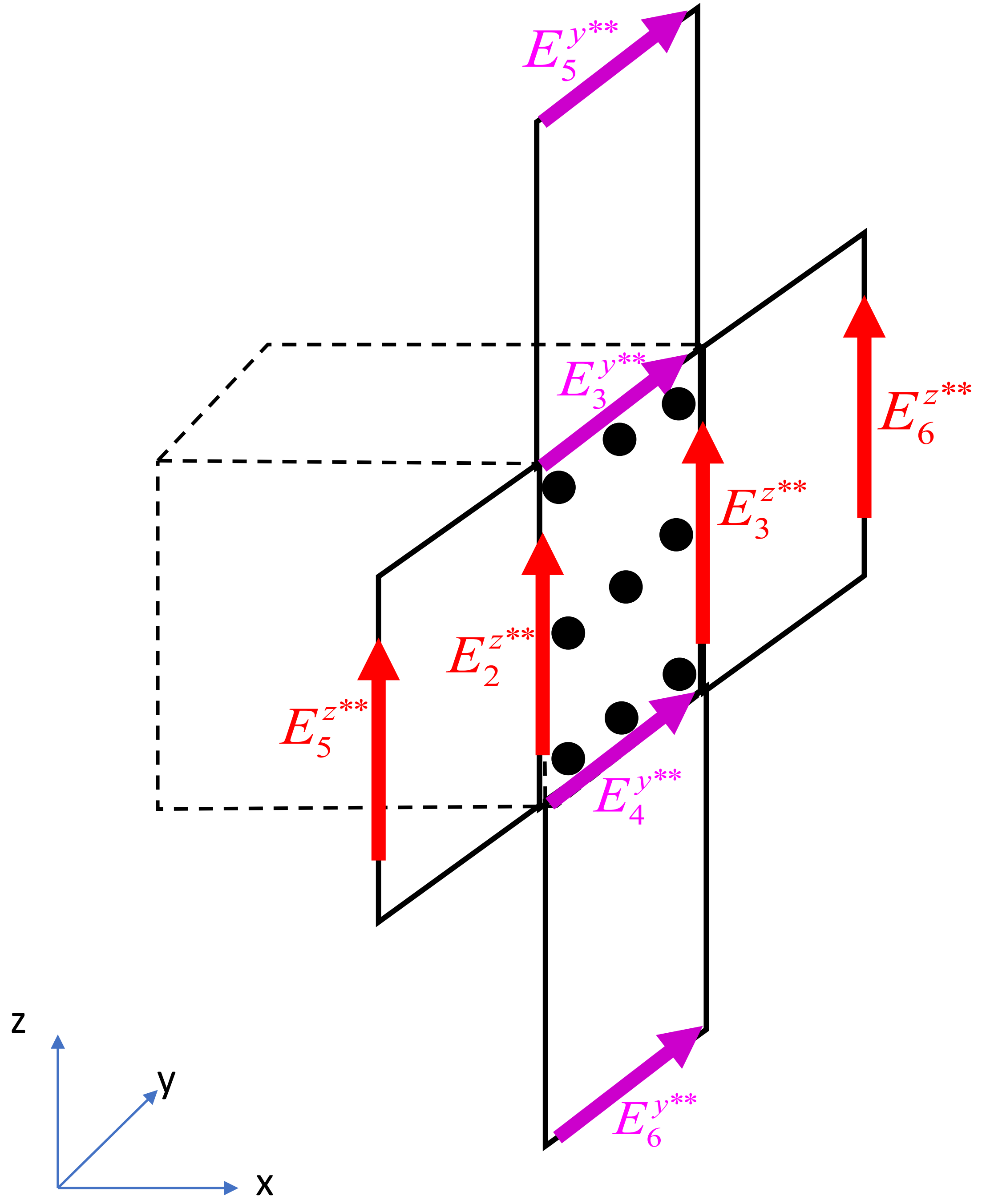

*Fig. 2 shows the right x-face of the zone shown in Fig. 1. The cube from Fig. 1 is now shown with dashed lines in this figure. In this figure, we show the electric fields not just in that x-face, but we also show relevant electric fields in the four x-faces that are adjacent to the face in question. All these edge-based electric fields are obtained from a two-dimensional Riemann solver, which is why we show them with a double star superscript. These electric fields, as well as their corresponding transverse magnetic fields, will be used for integrating the Poynting flux over the x-face that is shown in the figure. The nodal points within the x-face where the one-dimensional Riemann solvers are invoked for a higher order scheme (i.e. a third or fourth order scheme) are also shown as black dots. (For a second order scheme, the one-dimensional Riemann solver is invoked only at one node at the face center.)*

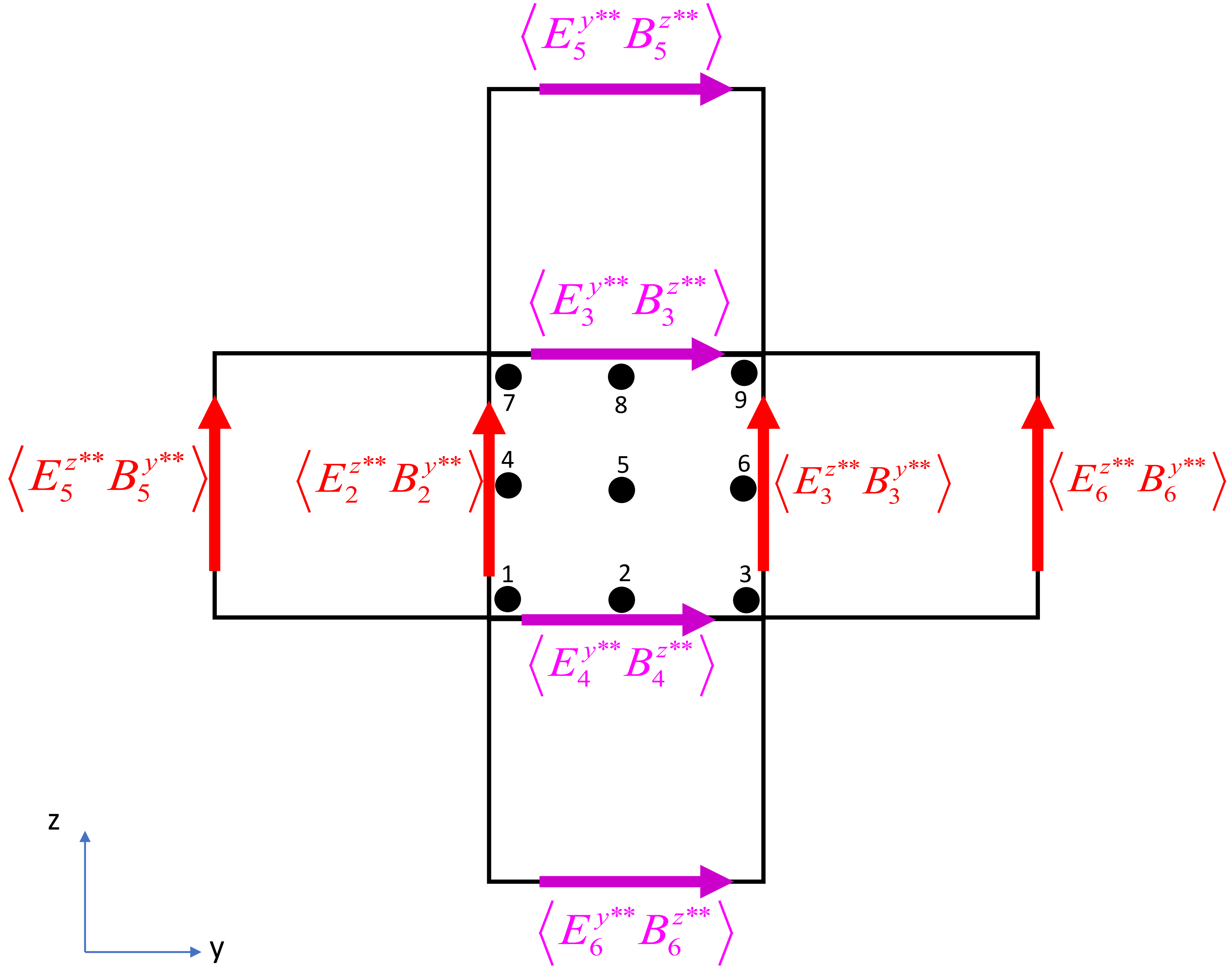

*Fig. 3 shows the same x-face that was highlighted in Fig. 2. But now we show all the nodal points in the x-face with numerically labeled dots. At these nodal points, one-dimensional Riemann solvers are invoked so that we can use the electric and magnetic fields from those Riemann solvers to construct a fourth order accurate face-averaged Poynting flux. However, recall that the red and magenta arrows also show the z-edge aligned and y-edge aligned electric fields that are obtained from a two-dimensional Riemann solver. The two-dimensional Riemann solver also yields the transverse magnetic fields. These have been used to obtain the angled brackets, which represent line integrals of the relevant products along each edge. In the text we show how these angled brackets are combined to obtain a fourth order accurate Poynting flux at the x-face that is exclusively based on the edge-centered information. This Poynting flux is more consistent with the update of the face-centered magnetic fields. It should, therefore, replace the Poynting flux that is based on the one-dimensional Riemann solvers.*

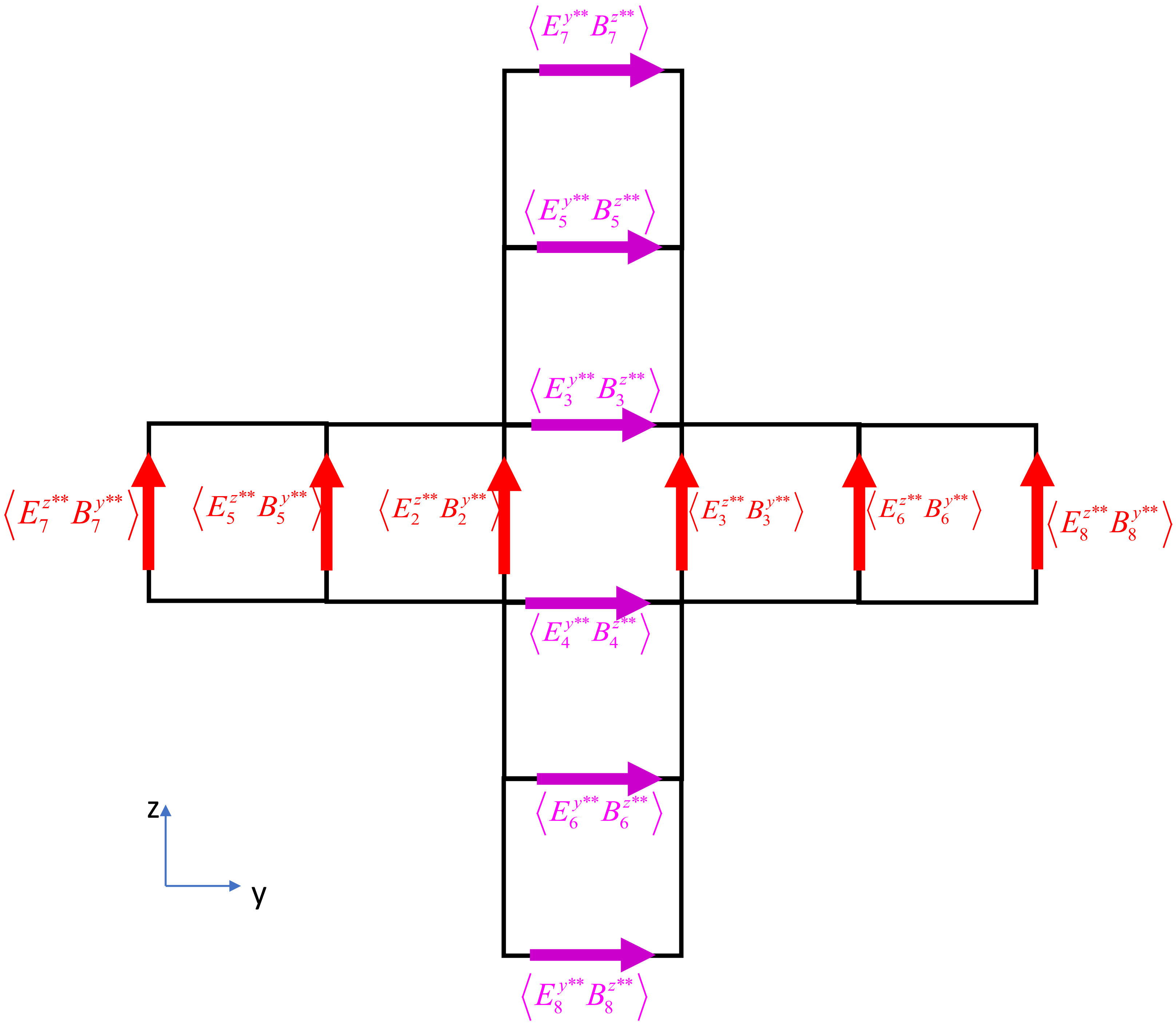

*Fig. 4 shows the same x-face that was highlighted in Fig. 3, but now we show a further layer of zones, suitable for up to sixth order of accuracy. We do not show nodal points within the face where one-dimensional Riemann solvers are invoked because it is very easy to envision a tensor product of nodal points for constructing a sixth order accurate face-averaged Poynting flux. However, the red and magenta arrows still show the z-edge aligned and y-edge aligned electric fields that are obtained from a two-dimensional Riemann solver. The two-dimensional Riemann solver also yields the transverse magnetic fields. These have been used to obtain the angled brackets, which represent line integrals of the relevant products along each edge. In the text we show how these angled brackets are combined to obtain a sixth order accurate Poynting flux at the x-face that is exclusively based on the edge-centered information. This Poynting flux is more consistent with the update of the face-centered magnetic fields. It should, therefore, replace the Poynting flux that is based on the one-dimensional Riemann solvers.*

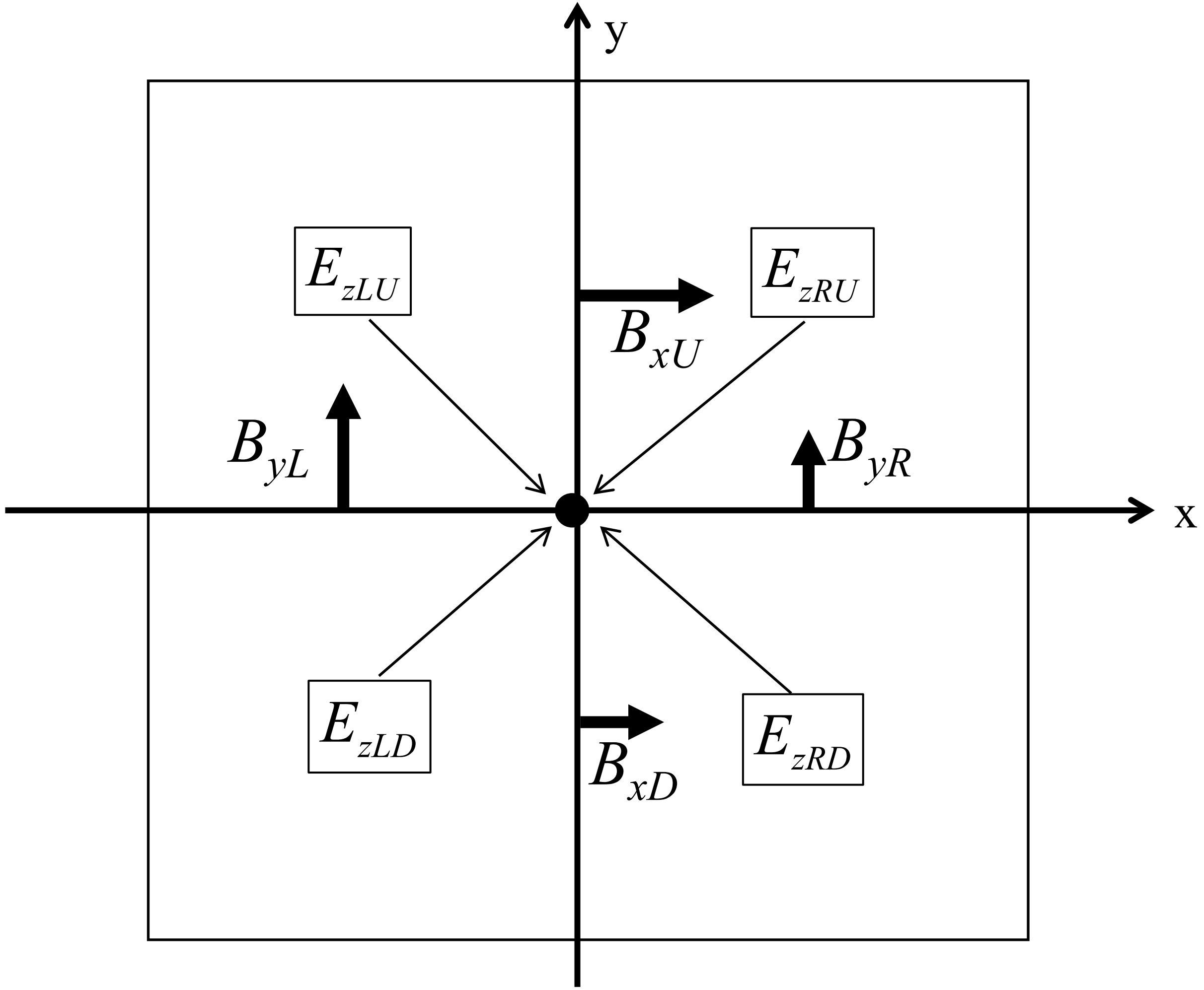

*Fig. 5 shows four zones in the xy-plane that come together at the z-edge of a three-dimensional mesh. Since the mesh is viewed from the top in plan view, the z-edge is shown by the black dot and the four abutting zones are shown as four squares. The four incoming states have subscripts given by "RU" for right-upper; "LU" for left-upper; "LD" for left-down and "RD" for right-down. The thin oblique arrows indicate that higher-order interpolation can eventually be used to obtain the centered part of the electric field at the z-edge. The thick horizontal and vertical arrows denote the normal components of the magnetic field at the zone faces. Owing to the divergence constraint, these field components are continuous across zone faces. They can, therefore, be obtained from the higher order facial reconstruction of the normal component of the magnetic field within each face. This provides higher order values of the x- and y-components of the magnetic field at the z-edge that minimize the dissipation terms.*

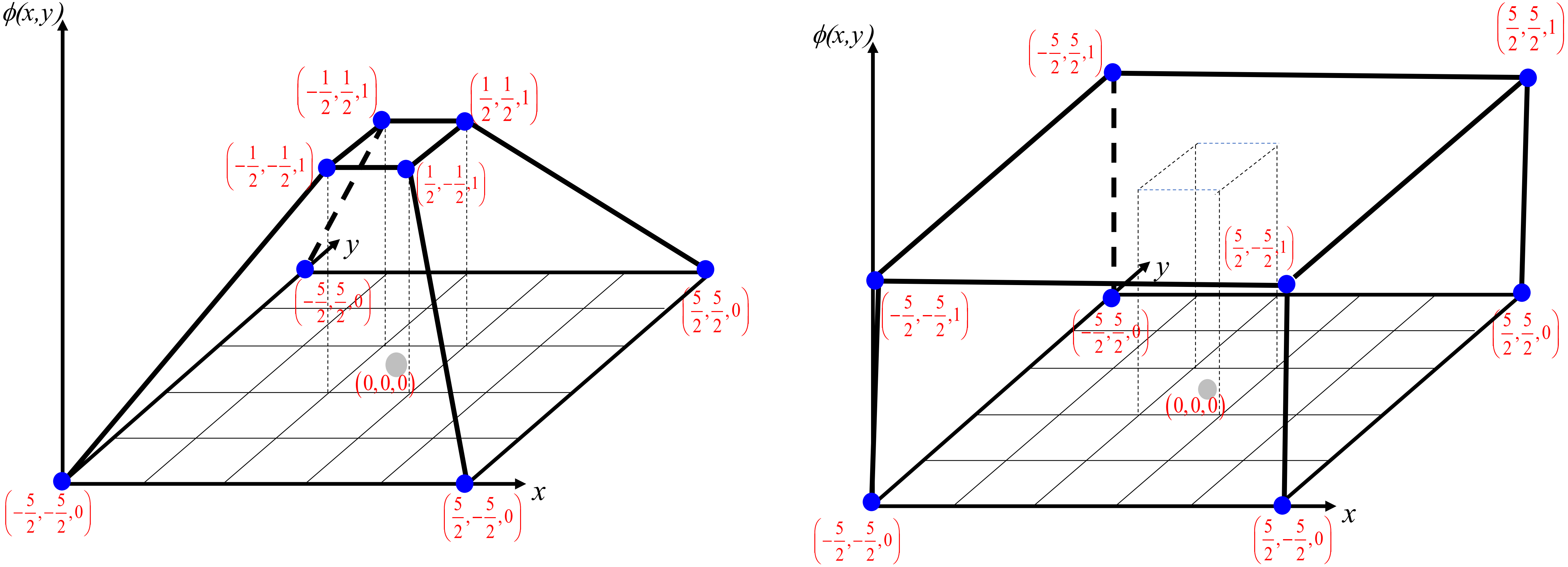

*Fig 6) Fig 6a Shows a two-dimensional tent function with a two zone taper in either direction. The central zone where the tent function is applied is shown by the gray dot and has local coordinates (0,0,0). The rest of the coordinates, shown in red, are relative to that origin. The end points of the tent function are shown by blue dots and have the coordinates shown in the figure. (Three-dimensional tent functions can be constructed similarly.) Fig. 6b shows a two-dimensional top-hat function with a two zone extent in either direction.*

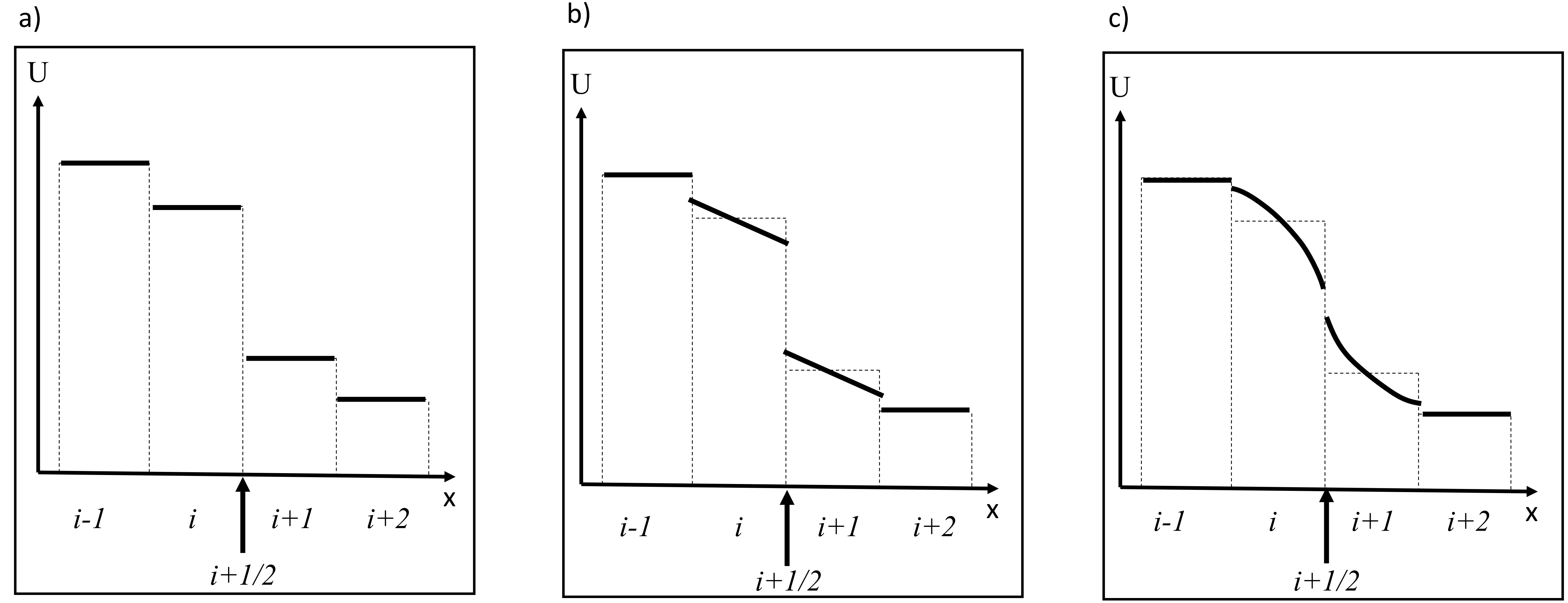

*Fig 7) shows three panels with the same mesh function in one dimension for a sub-section of a mesh with four adjoining zones. The mesh function is shown by the dashed lines in all the panels. The thick solid line in Fig. 7a shows a first order reconstruction. The thick solid line in Fig. 7b shows a second order reconstruction. The thick solid line in Fig. 7c shows a third, or higher, order reconstruction. Our focus is on the zone boundary, shown by the arrow. We see that with increasing order, the jump in the reconstructed function at the zone boundary of interest decreases.*

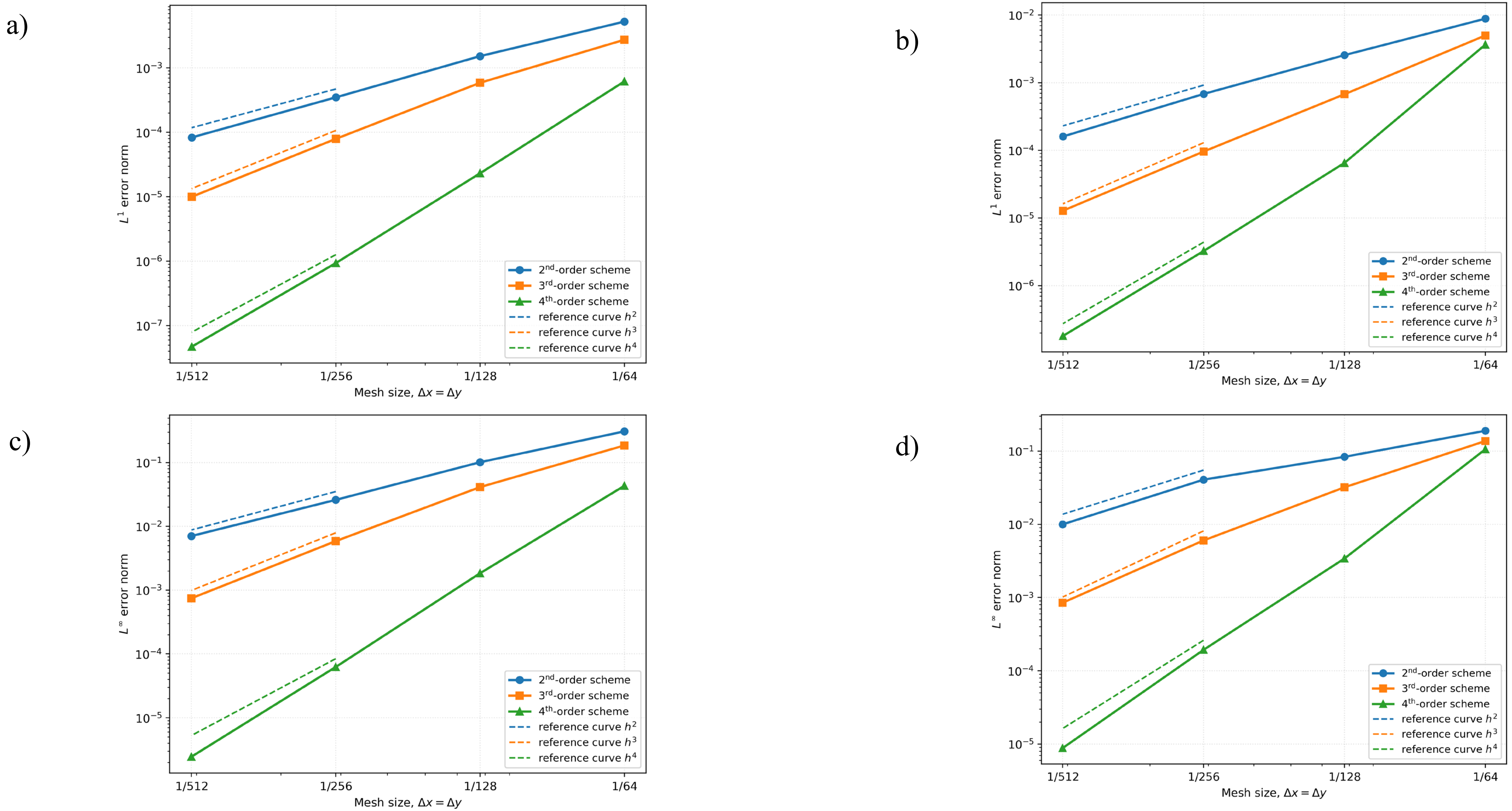

*Fig. 8) Error analysis of the Vortex problem for the MHD and RMHD System: Fig 8a shows the $L^1$-error plots for the non-relativistic magnetized vortex, and Fig 8b shows the $L^1$-error plots for the relativistic magnetized vortex. Fig 8c shows the $L^{inf}$-error plots for the non-relativistic magnetized vortex, and Fig 8d shows the $L^{inf}$-error plots for the relativistic magnetized vortex. These errors are shown for the $B_y$ variable (y-component of the magnetic field). The error slopes match the reference slopes at higher resolution, showing that the expected convergence rate is recovered.*

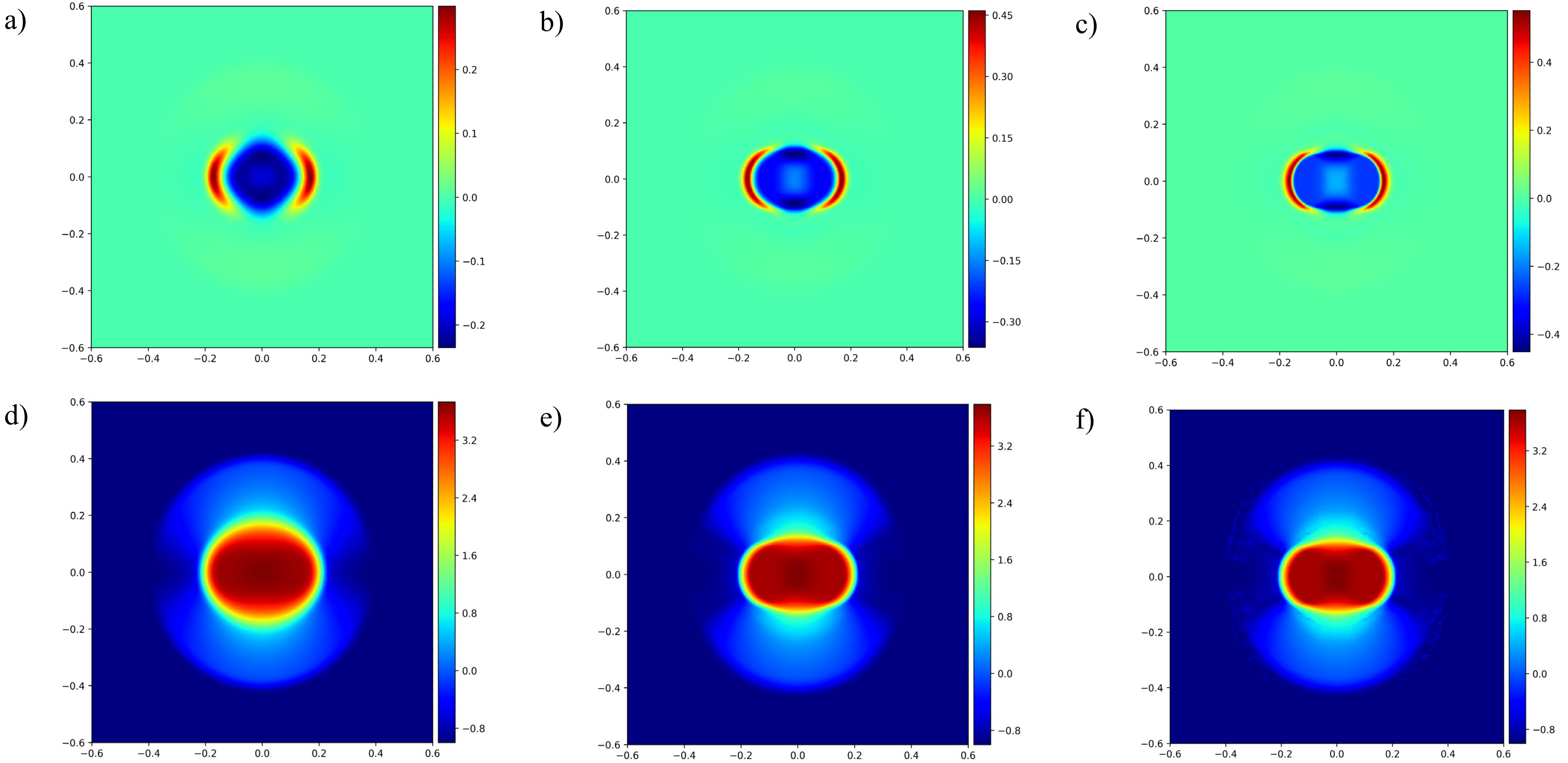

*Fig. 9) MHD Blast Problem: Fig. 9a, 9b and 9c (top panel) show the log (base 10) of the density profile when using a 1st, 2nd and 4th order scheme, respectively. Fig. 9d, 9e and 9f (bottom panel) show the log (base 10) of the pressure profile when using a 1st, 2nd and 4th order scheme, respectively. All the profiles were obtained at time t=0.001 on a two-dimensional grid that consists of 400×400 zones. The pressure ratio between the initial blast pressure and the ambient pressure is 10,000:0.1. The initial plasma beta in the ambient medium for this problem is $\beta$=2.51×10$^{-6}$.*

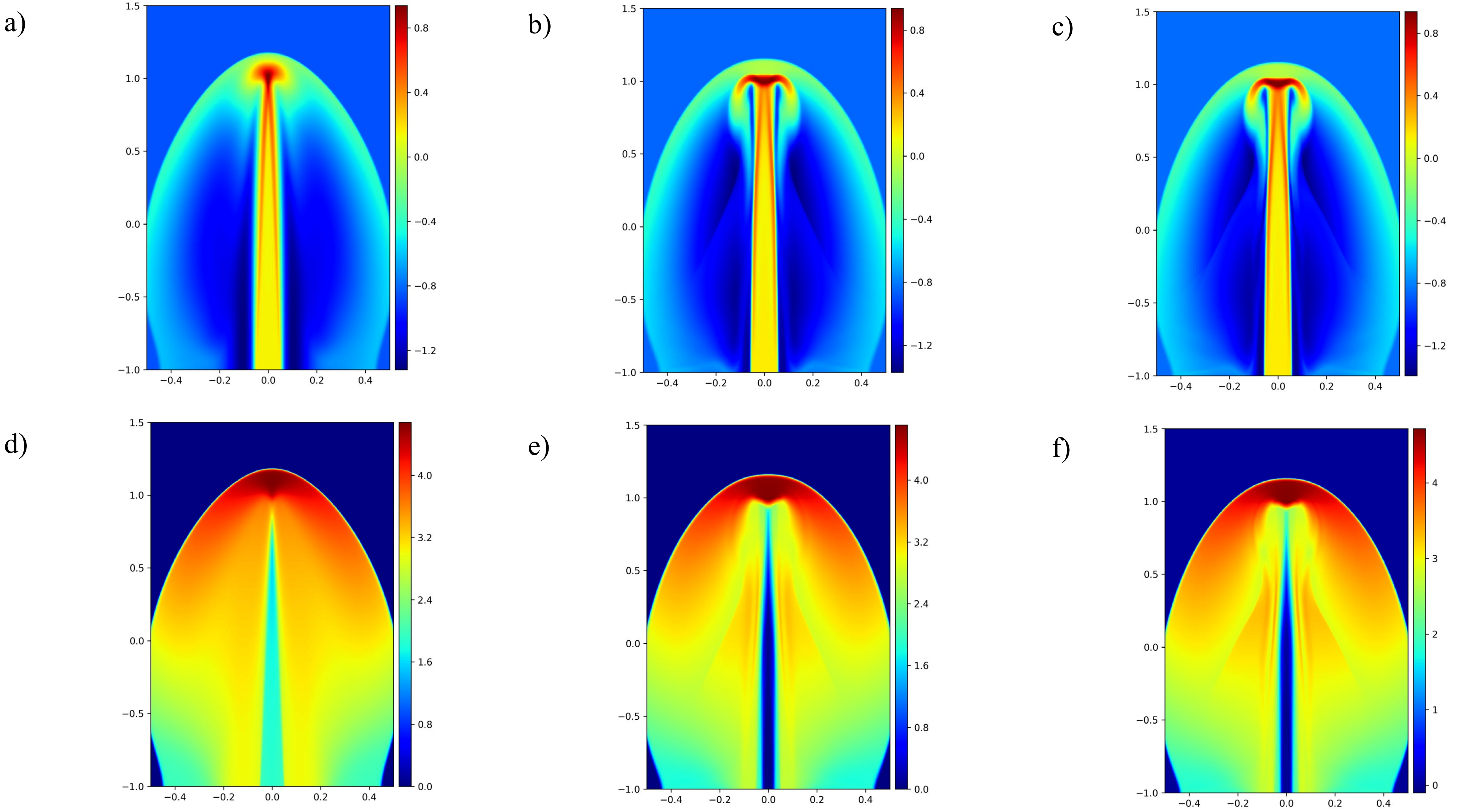

*Fig. 10) MHD Jet Problem: Fig. 10a, 10b and 10c (top panel) show the logarithm of density (base 10) when using a 1st, 2nd and 4th order scheme, respectively. Fig. 10d, 10e and 10f (bottom panel) show the logarithm of pressure (base 10) when using a 1st, 2nd, and 4th order scheme, respectively. All the profiles were obtained at t=0.002 on a two-dimensional grid consisting of 400×600 zones. The inlet Mach number of the jet is 800. The initial plasma beta for this problem is* $\beta = 10^{-4}$ .

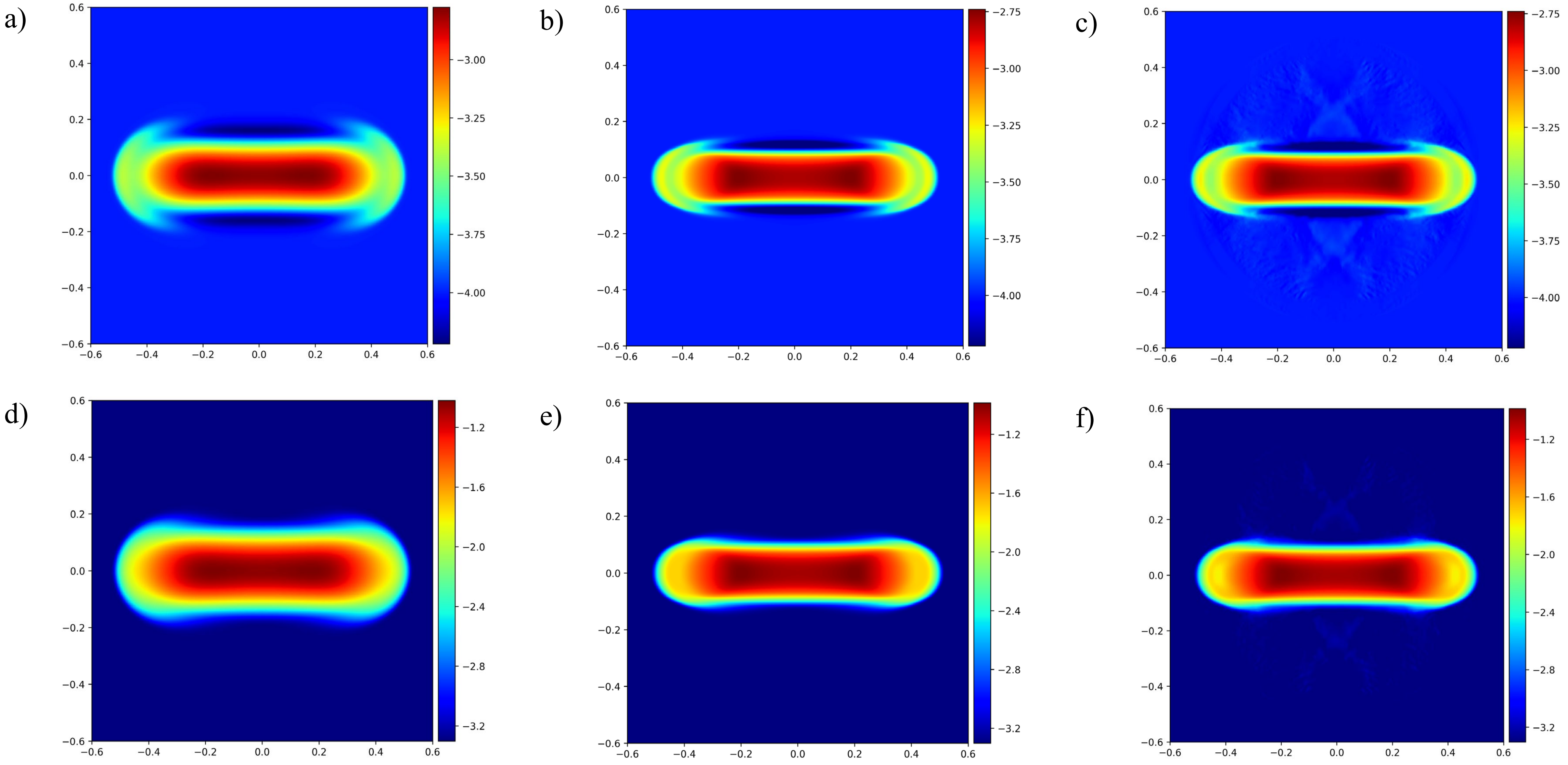

*Fig. 11) RMHD Blast Problem: Fig. 11a, 11b and 11c (top panel) show the log (base 10) of the density profile when using a 1st, 2nd and 4th order scheme, respectively. Fig. 11d, 11e and 11f (bottom panel) show the log (base 10) of the pressure profile when using a 1st, 2nd and 4th order scheme, respectively. All the profiles were obtained at time t=4 on a two-dimensional grid that consists of 400×400 zones. The initial pressure for this problem was 2.5×10⁻⁶ times smaller than the magnetic pressure (B²/2).*

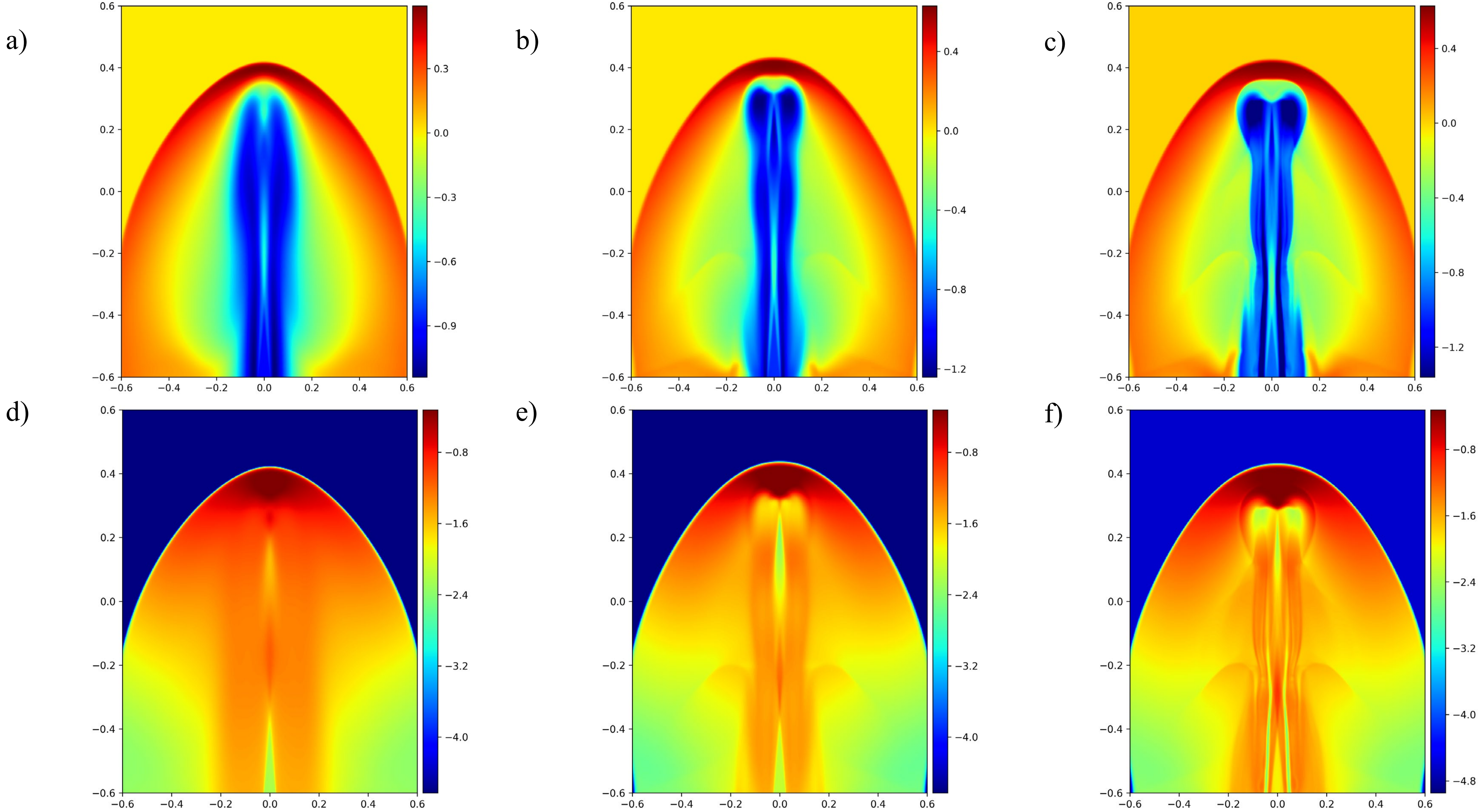

*Fig. 12) RMHD Jet Problem: Fig. 12a, 12b and 12c (top panel) show the logarithm of density (base 10) when using a 1st, 2nd and 4th order scheme, respectively. Fig. 12d, 12e and 12f (bottom panel) show the logarithm of pressure (base 10) when using a 1st, 2nd, and 4th order scheme, respectively. All profiles were obtained at t=30.0 on a two-dimensional grid with 480×500 zones. The inlet Lorentz factor for the jet is 7.09. The initial pressure for this problem was $10^{-3}$ times smaller than the magnetic pressure ($B^2/2$).*

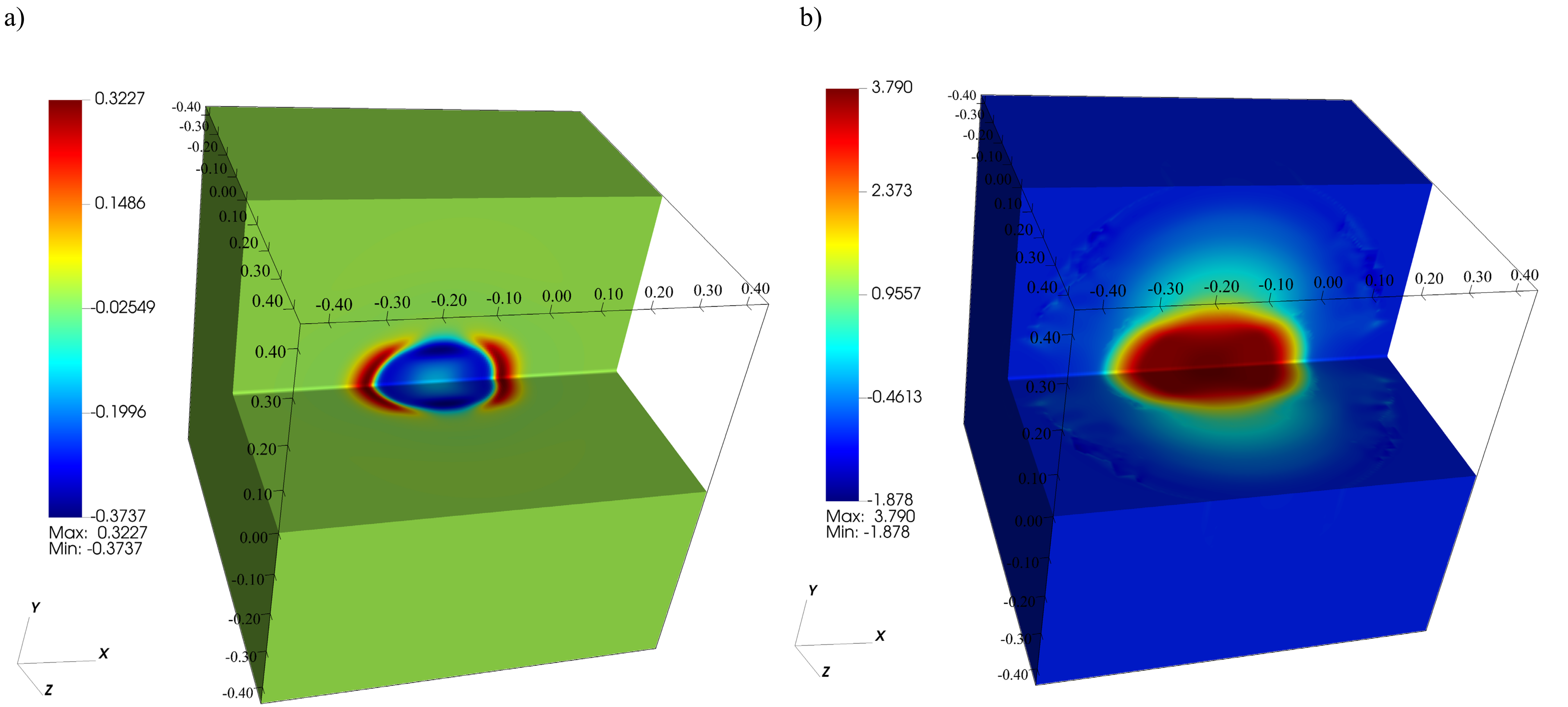

*Fig. 13) MHD Three-Dimensional Blast Problem: Fig. 13a shows the logarithm (base 10) of the density profile and Fig. 13b shows the logarithm (base 10) of the pressure profile for the three-dimensional MHD-Blast at time t=0.001 on a three-dimensional grid that consists of 200×200×200 zones. The pressure ratio between the initial blast pressure and the ambient pressure is 10,000:0.1. The initial plasma beta in the ambient medium for this problem is* $\beta=2.51\times10^{-6}$.

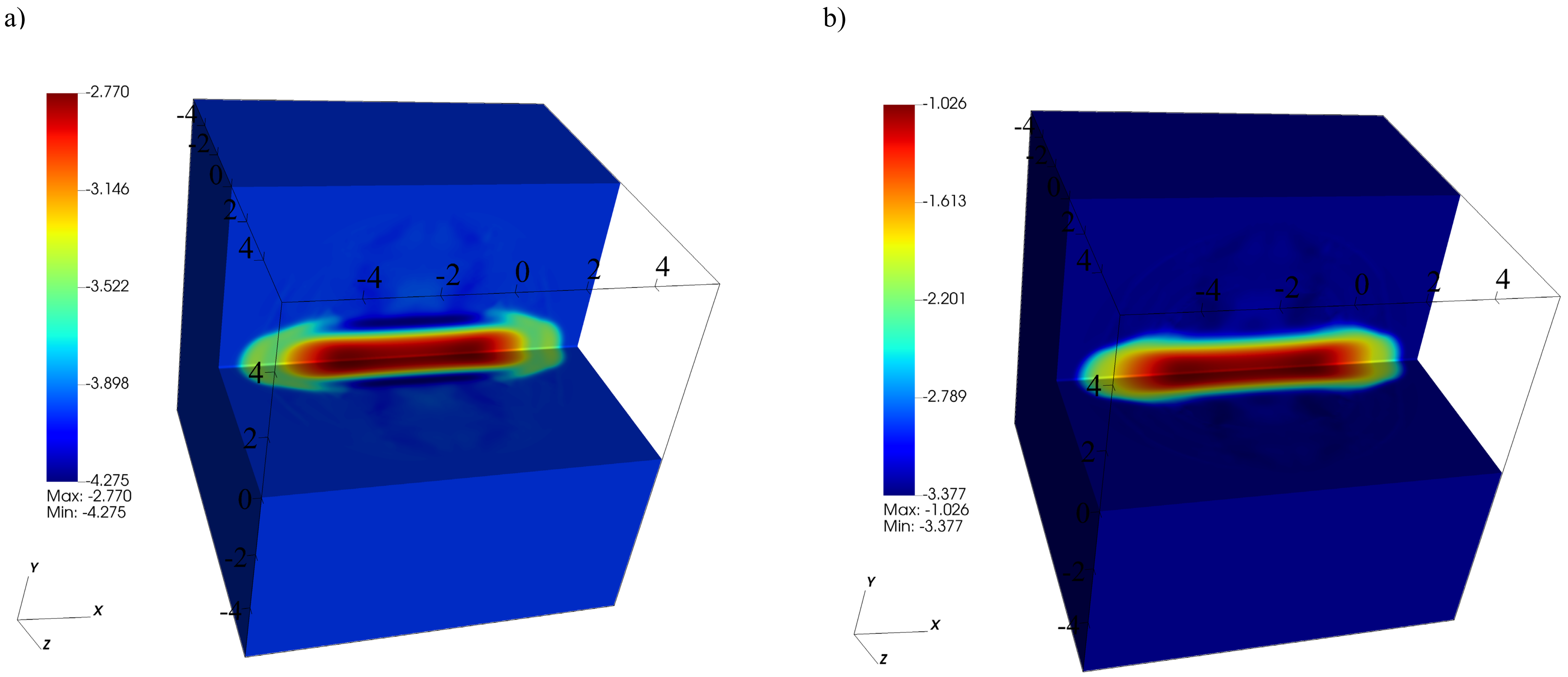

*Fig. 14) Three-Dimensional Relativistic MHD Blast Problem: Fig. 14a shows the log (base 10) of the density profile and Fig. 14b shows the log (base 10) of the pressure profile for the three-dimensional RMHD-Blast at time t=4.0 on a three-dimensional grid that consists of 200×200×200 zones. The initial pressure for this problem was $2.5\times10^{-6}$ times smaller than the magnetic pressure ($B^2/2$)*